\documentclass[10pt]{article}
\usepackage[margin=1in]{geometry}
\usepackage{microtype}
\usepackage{amsmath,amssymb,amsfonts,amsthm,mathtools}
\allowdisplaybreaks[1]
\usepackage{xcolor}
\usepackage{color}
\usepackage{float}
\usepackage{placeins}
\usepackage{booktabs}
\usepackage{graphicx}
\usepackage[compatibility=false]{caption}
\usepackage{subcaption}
\usepackage{tikz}
\usepackage[colorlinks=true,linkcolor=blue,citecolor=blue,urlcolor=blue]{hyperref}

\newcommand{\SO}{\mathcal{SO}}
\newcommand{\R}{\mathbb{R}}
\newcommand{\E}{\mathbb{E}}

\newcommand{\C}{\mathbb{C}}
\newcommand{\cP}{\mathcal{P}}

\newcommand{\cT}{\mathcal{T}}
 
\title{\textit{Ab-initio} cryo-EM  reconstruction from moments of denoised images}

\author{
Yunpeng Shi
    \thanks{Department of Mathematics, University of California, Davis, ({\tt ypshi@ucdavis.edu}).} 
\quad
Amit Singer
	\thanks{Department of Mathematics and Program in Applied and Computational Mathematics, Princeton University, 
	({\tt amits@math.princeton.edu}).} 
\quad
Yuguan Wang
\thanks{Corresponding author. Program in Applied and Computational Mathematics, Princeton University, ({\tt yuguanw@princeton.edu}).
         }
 \quad 
Andy Zhang 
\thanks{Program in Applied and Computational Mathematics, Princeton University, ({\tt zhang.andy@princeton.edu}).
         }
	 }
\date{\today}

\newcommand{\keywords}[1]{%
  \par\medskip
  \noindent\textbf{Keywords:} #1%
}

\begin{document}
\maketitle

\begin{abstract}
In single-particle cryo-electron microscopy (cryo-EM), high-resolution three-dimensional (3-D) refinement requires a reliable \textit{ab-initio} model to converge to the correct structure. Both refinement and existing \textit{ab-initio} methods estimate the orientations of the particle images together with the 3-D structure, which can be slow for large and noisy datasets.  We present a method of moments approach that bypasses orientation estimation and recovers a 3-D \textit{ab-initio} model from the first three moments of denoised images. On experimental  datasets, our method produces \textit{ab-initio} models rapidly and successfully initializes high-resolution refinement.  This is the first successful application of  method-of-moments-based  \textit{ab-initio} modeling to experimental datasets without assuming a uniform distribution of orientations.

\keywords{cryo-EM, \textit{ab-initio} modeling, method of moments, image denoising.}
\end{abstract}

\section{Introduction}
Single-particle cryo-EM is a powerful technique for determining the 3-D structures of biological macromolecules at high resolution~\cite{shen2018nobel}, and even atomic resolution~\cite{Nakane2020,Yip2020}. In a cryo-EM experiment, many copies of a molecule are frozen in a thin layer of ice and imaged with an electron microscope. The microscope produces large images, called micrographs, each containing many noisy two-dimensional (2-D) tomographic projections of randomly rotated molecules. These projections are  extracted from the micrographs by a procedure called particle picking~\cite{bepler2019positive}, which yields the particle images used for reconstruction.  To limit radiation damage, the images are acquired at low electron doses and are therefore extremely noisy. 

In the most popular software packages, such as RELION~\cite{scheres2012relion} and cryoSPARC~\cite{punjani2017cryosparc}, high-resolution reconstruction is based on maximum likelihood estimation, in which the unknown orientations and translations are treated as latent variables. The likelihood is maximized by expectation--maximization (EM)~\cite{dempster1977maximum}, a procedure known as iterative refinement. In each iteration, every image is aligned against the current structure by computing the posterior distribution of its orientation and translation, and the structure is then reconstructed from the images weighted by these posteriors. Iterative refinement has been remarkably successful, but it is computationally expensive, since every image is compared with projections over many orientations and translations in each iteration. Moreover, since the likelihood is highly nonconvex, iterative refinement may converge to a poor local maximum, and therefore requires a good \textit{ab-initio} model for initialization.

A widely used \textit{ab-initio}  method is to maximize the likelihood using stochastic optimization~\cite{punjani2017cryosparc,10.1042/BCJ20210708}. It starts from a random initial structure and updates it using gradients computed on random subsets of images. A fundamentally different approach is based on \textit{common lines}~\cite{VANHEEL1987111,Vainshtein1986DeterminationOT,Bandeira_2020,singer2010detecting,singer2011three,shkolnisky2012viewing,GREENBERG2017106}. The Fourier transforms of any two projection images of the same molecule share a common line along which they agree. The common lines between pairs of images or 2-D class averages~\cite{ZHAO2014153} can be used to determine their relative rotations, from which the 3-D structure is reconstructed. All of the above approaches need to estimate the orientations, either explicitly or through their posterior distributions. This is computationally expensive for large datasets, and at high noise levels the rotations cannot be estimated reliably, especially for small molecules. It is therefore desirable to have a fundamentally different approach that avoids orientation estimation altogether.

Kam's method, also known as the method of moments, was proposed by Zvi Kam in 1980~\cite{kam1980reconstruction}. The main idea is that low-order statistics of the images (such as the mean and covariance) can be reliably estimated from the noisy images by averaging and bias subtraction. Since these statistics are determined by the structure, one first computes the sample moments from the images, and then recovers the structure directly from the moments without pose estimation. Only one or a few passes over the data are needed, so the method of moments is computationally efficient on large datasets and robust to noise. Moreover, it enjoys provable guarantees under certain conditions~\cite{sharon2020method,bendory2023autocorrelation,kileel2025two,balanov2026provable,BANDEIRA2023236,doi:10.1137/23M155685X}. However, Kam's method was largely  overlooked because the datasets at that time were too small to accurately estimate the moments. It also assumes that the orientations are uniformly distributed, which does not hold in practice~\cite{tan2017addressing}. In contrast, Kam's method has been actively developed for single-particle reconstruction with X-ray free-electron lasers (XFELs)~\cite{kirian2012structure,donatelli2015iterative,kurta2017correlations,pande2018ab,zhao2024structure}, where the particle orientations are uniformly distributed.

Over the past decade, a series of efforts have been made to apply Kam's method to experimental datasets. The first work by Bhamre, Zhang and Singer~\cite{bhamre2015orthogonal,Bhamre2017AnisotropicTF} assumed that the orientations are uniformly distributed and used only the second moment. The resulting moment-matching problem is very ill-posed and requires two denoised reference images to resolve the ambiguity~\cite{levin20183d}. Later work in the same setting resolved the ambiguity by promoting sparsity~\cite{bendory2023autocorrelation,doi:10.1137/22M1498218}. To handle non-uniform distributions of orientations, Sharon et al.~\cite{sharon2020method} showed that, when the structure and the orientation distribution are expanded in steerable bases, matching the first two moments yields polynomial equations in the expansion coefficients, which can then be solved using gradient-based optimization. Recent work~\cite{doi:10.1137/24M1699644} incorporated a compressed representation of the third moment, which alleviates the ill-posedness of using only the first two moments. Another recent work~\cite{kileel2025two} showed that the second moments of two datasets, one with a uniform distribution of viewing directions and the other with a non-uniform distribution of viewing directions, together determine the structure without the third moment.  These methods perform well on synthetic datasets, where the images exactly follow the assumed image model. However, reliable estimation of moments from raw experimental images remains challenging, particularly for the third moment, which is sensitive to any mismatch between the data and the image model. Experimental datasets may contain neighboring particles, which present a significant challenge for moment estimation~\cite{andy2026overcrowded}. The images may also exhibit colored noise, varying contrast, and residual background offsets. These effects are not fully accounted for in existing formulations.

\begin{figure}[ht!]
\centering
\includegraphics[width=\textwidth]{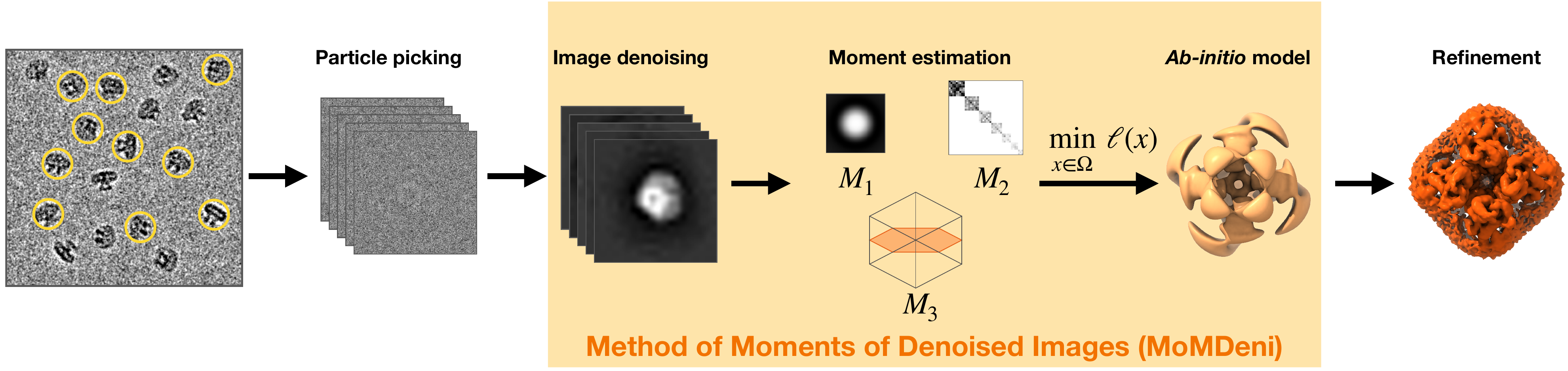}
\caption{Overview: MoMDeni denoises the picked particles and uses them to estimate the moments, followed by reconstructing an \textit{ab-initio} model from the moments through numerical optimization.}
\label{fig:overview}
\end{figure}

In this work, we propose the method of moments of denoised images (MoMDeni), which uses image denoising as a practical preprocessing step for moment-based reconstruction. We apply covariance Wiener filtering~\cite{bhamre2016denoising,shi2022contrast} to suppress noise and correct for the contrast transfer fuctions (CTFs). The moments of the denoised images can then be computed directly. The denoised images also make neighboring particles easier to identify, allowing us to discard overcrowded images before computing the moments. We then use the first three moments of the selected denoised images to recover the 3-D structure within the method-of-moments framework for non-uniform orientation distributions~\cite{sharon2020method}. Figure~\ref{fig:overview} gives a schematic overview of MoMDeni. Denoising changes the image statistics, so these moments are generally biased estimates of the moments of the clean projections. Nevertheless, our experiments show that when the denoised images have enough structural information, these moments still yield useful \textit{ab-initio} models. On six experimental datasets, we demonstrate that MoMDeni can successfully initialize iterative refinement. Our code is publicly available at \href{https://github.com/wangyuguan/MoMDeni}{https://github.com/wangyuguan/MoMDeni}.

\section{Preliminaries}
In this section, we introduce the mathematical notation and results needed to describe our method. We first introduce the notation in Section~\ref{sec:notations}. We then describe the cryo-EM image model in Section~\ref{sec:image_model}. Finally, we introduce the function bases used to build the method of moments framework in Section~\ref{sec:steerable}.

\subsection{Notation}
\label{sec:notations}
For a function $f:\R^d\to \R$, we denote its $d$-dimensional Fourier transform by
\begin{equation}
    \mathcal F[f](\xi)=\int_{\R^d} e^{-2\pi i x\cdot \xi} f(x)\,dx.
\end{equation}
We also define the inverse Fourier transform
\begin{equation}
    \mathcal F^{-1}[f](x)=\int_{\R^d} e^{2\pi i x\cdot \xi} f(\xi)\,d\xi.
\end{equation}
A function is defined to be \textit{bandlimited} with a cut off frequency $c>0$ if its Fourier transform is compactly supported in the ball $B_c(0)$. In this work, we choose $c=1/2.$

Let $*$ denote convolution such that 
\begin{equation}
    (f*g)(x)=\int_{\R^d} f(x-y) g(y)\,dy.
\end{equation}
The Fourier transform of the convolution is given by 
\begin{equation}
    \mathcal{F}[f*g](\xi)=\mathcal{F}[f](\xi)\mathcal{F}[g](\xi).
    \label{eqn:convolution}
\end{equation}

For $\tau\in\R^d$, we define the shift operator
\begin{equation}
    \cT_\tau f(x) :=f(x-\tau).
\end{equation}
In the Fourier domain, it becomes a phase modulation,
\begin{equation}
    \mathcal{F}[\cT_\tau f](\xi)=e^{-2\pi i\,\xi\cdot\tau}\,\mathcal{F}[f](\xi).
    \label{eqn:shift}
\end{equation}

We denote the group of 3-D rotations by
\begin{equation}
    \SO(3)=\{R\in\R^{3\times 3} \,|\, RR^T=I_{3\times 3},\ \det(R)=1\}
\end{equation}
where $I_{d\times d}$ stands for the identity matrix of dimension $d$.

We define the tomographic projection of a 3-D function along the direction determined by $R$ as 
\begin{equation}
 ( \cP_R f) (x_1,x_2):=\int_\R f(R^T(x_1,x_2,x_3)^T) \,dx_3.
\end{equation}
By the Fourier-slice theorem, the 2-D Fourier transform of $ \cP_R f$ is given by the central slice
\begin{equation}
    \mathcal{F}[\cP_R f](\xi_1,\xi_2)= \mathcal{F}[f](R^T(\xi_1,\xi_2,0)^T). 
    \label{eqn:fourier-slice}
\end{equation}

Throughout, $\delta_{a,b}$ (also written $\delta_{ab}$) denotes the Kronecker delta, which is $1$ if $a=b$ and $0$ otherwise.

\subsection{Cryo-EM image model}
\label{sec:image_model}
Let $\Phi:\R^3 \to \R$ denote the density map of electric potential of a biomolecule, which we refer to as the \textit{volume}. Under a standard cryo-EM image model, the particle images are generated according to 
\begin{equation}
    I_j(x_1,x_2)= \alpha_j \mathcal{F}^{-1}[H_j]*(\cT_{\tau_j}\cP_{R_j} \Phi)(x_1,x_2)+\epsilon_j(x_1,x_2),\quad j=1,\ldots,N,
\label{eqn:image_model}
\end{equation} 
where $R_j$ is a random unknown rotation, $\tau_j\in \R^2$ is an unknown offset caused by the inaccuracy of particle picking, $\alpha_j>0$ is a random amplitude contrast, $\epsilon_j$ is random noise in image $j$  whose standard deviation depends on $j$, and $H_j$ is the CTF. The CTFs can be estimated using standard tools such as CTFFIND4~\cite{ROHOU2015216}, so we assume that they are provided.
In practice, the images are observed on a $K\times K$ pixel grid. We keep the continuous notation for readability.

By equations~\eqref{eqn:convolution}, \eqref{eqn:shift} and \eqref{eqn:fourier-slice}, the Fourier transform of the $j$-th image is
\begin{equation}
    \mathcal F[I_j](\xi)=\alpha_j\,H_j(\xi)\,e^{-2\pi i\,\xi\cdot \tau_j}\,\mathcal F[\Phi]\big(R_j^T(\xi_1,\xi_2,0)^T\big)
    +\mathcal F[\epsilon_j](\xi),
\end{equation}
where $\xi=(\xi_1,\xi_2)$. In practice, the noise is colored. We estimate its radial power spectrum from the margins of all images together, rather than separately for each micrograph, because some micrographs contain too few images. We then whiten the images, which only changes each CTF $H_j$ by a radial factor. We therefore assume from now on that $\epsilon_j$ is white Gaussian noise and $H_j$ denotes the whitened CTF.

\subsection{Basis representation}
\label{sec:steerable}
We introduce convenient basis representations used in our method of moments.  

\subsubsection{Image representation}
We assume the image is bandlimited by $c>0$ and represent its Fourier transform in a 2-D steerable basis with expansion order $Q>0$
\begin{equation}
    \mathcal{F}[I](r,\varphi)=\sum_{q=-Q}^Q\sum_{t=1}^{T_q} a_{q,t} f_{q,t}(r) e^{iq\varphi}
    \label{eqn:image_basis}
\end{equation}
where $a_{q,t} \in \C$ are the coefficients and $f_{q,t}$ are real radial basis functions with $f_{-q,t}=f_{q,t}$, satisfying 
\begin{equation}
    \int_0^{c}f_{q,t}(r)f_{q,t'}(r) r\,dr=\delta_{tt'}.
\end{equation}
Because the images $I_j$ are real, we have the constraint 
\begin{equation}
    a_{-q,t}=(-1)^q \overline{a_{q,t}}. 
\end{equation}

\subsubsection{Volume representation}
We assume the volume is bandlimited by $c>0$ and represent its Fourier transform in a 3-D steerable basis with expansion order $L>0$
\begin{equation}
   \mathcal F [\Phi_A](r,\theta,\varphi)=\sum_{l=0}^L\sum_{s=1}^{S_l}\sum_{m=-l}^l A_{l,m,s} F_{l,s}(r) Y_l^m(\theta,\varphi),
   \label{eqn:sfb3d}
\end{equation}
where $(r,\theta,\varphi)$ are the spherical coordinates of the frequency with $r\in[0,c]$, and $\Phi_A$ denotes the volume determined by the expansion coefficients $A=\{A_{l,m,s}\}$. Each basis function is the product of a radial part $F_{l,s}$ and an angular part $Y_l^m$.

For each $l$, the radial parts $F_{l,s}$, $s=1,\dots,S_l$, are real functions satisfying
\begin{equation}
    \int_0^c  F_{l,s}(r) F_{l,s'}(r) r^2\,dr = \delta_{ss'}.
\end{equation}
For the angular parts, let $\mathbb{S}^2$ be the unit sphere in $\R^3$. We identify a point $\eta\in\mathbb{S}^2$ with its spherical coordinates $(\theta,\varphi)$, where $\theta\in[0,\pi]$ is the polar angle and $\varphi\in[0,2\pi)$ is the azimuthal angle. The functions $Y_l^m:\mathbb{S}^2 \to \C$ are the spherical harmonics
\begin{equation}
    Y_l^m(\theta,\varphi)=(-1)^{m} \sqrt{\frac{2l+1}{4\pi} \frac{(l-|m|)!}{(l+|m|)!} }  P_l^{|m|}(\cos\theta) e^{im\varphi}
    \label{eqn:sph_harm}
\end{equation}
with $l\ge 0$ and $-l\le m \le l$, where $P_l^m$ are the associated Legendre functions~\cite{abramowitz1964handbook}. With this convention, $Y_l^{-m}=\overline{Y_l^m}$.

Because the volume $\Phi_A$ is real, its Fourier transform satisfies $\mathcal F[\Phi_A](-\xi)=\overline{\mathcal F[\Phi_A](\xi)}$. Since $Y_l^m(-\eta)=(-1)^lY_l^m(\eta)$, this gives the constraint
\begin{equation}
    A_{l,-m,s}=(-1)^l\,\overline{A_{l,m,s}}.
\end{equation}

 If the volume has symmetry, additional constraints can be imposed~\cite{Meyer_1954}.  For example,  we have
\begin{equation}
    A_{l,m,s}=0,\quad \text{if } m\not\equiv 0 \pmod{n}.
    \label{eqn:Cn}
\end{equation}
for a volume with $\mathbf{C}_n$ symmetry. 
For $\mathbf{D}_n$ symmetry, in addition to~\eqref{eqn:Cn}, we have
\begin{equation}
    A_{l,-m,s}=(-1)^{l+m} A_{l,m,s}.
    \label{eqn:Dn}
\end{equation}

For a volume $\Phi_A$ represented by~\eqref{eqn:sfb3d}, a rotation $R$ and a shift $\tau=(\rho\cos\psi,\rho\sin\psi)\in\R^2$, the Fourier transform of the shifted projection has the steerable expansion
\begin{equation}
    \mathcal F[\cT_\tau\cP_R\Phi_A](r,\varphi)
    =\sum_{q=-Q}^Q\sum_{t=1}^{T_q} a_{q,t}(A,R,\tau) f_{q,t}(r) e^{iq\varphi}.
    \label{eqn:proj_basis}
\end{equation}
where  $ a_{q,t}(A,R,\tau)$ are the expansion coefficients associated with the 3-D structure $A$, the orientation $R$ and the shift $\tau$. For $\tau=0$, the coefficients are given by~\cite{sharon2020method}
\begin{equation}
    a_{q,t}(A,R,0)=\sum_{l=|q|}^{L}\sum_{s=1}^{S_l}\sum_{m=-l}^{l} A_{l,m,s}\, D^l_{m,q}(R)\, \gamma^{q,t}_{l,s},
    \label{eqn:coef2d}
\end{equation}
where $D^l$ is the Wigner-D matrix of degree $l$~\cite{chirikjian2016harmonic}, and the constants are
\begin{equation}
    \gamma^{q,t}_{l,s}= Y_l^q(\tfrac{\pi}{2},0)\int_0^c F_{l,s}(r)\, f_{q,t}(r)\, r\,dr .
\end{equation}
 Since $P_l^{|q|}(0)=0$ when $l+q$ is odd, only degrees $l$ satisfying $l\equiv q \pmod 2$ contribute, and $a_{q,t}(A,R,0)=0$ for $|q|>L$. For this reason, we choose $Q\le L$. The radial functions $f_{q,t}$ used for the moments are described in Appendix~\ref{app:radial}.

\subsubsection{Orientation distribution representation}
Let $\mu$ be the density of the rotations $R_j$ with respect to the Haar measure $dR$~\cite{chirikjian2016harmonic} on $\SO(3)$. A rotation $R$ projects the volume along the viewing direction $R^Te_3$, where $e_3=(0,0,1)^T$. The rotations $R_z(\gamma)R$, $\gamma\in[0,2\pi)$, where $R_z(\gamma)$ is the rotation about the $z$-axis in~\eqref{eqn:rotz}, share the same viewing direction, and their projections differ only by an in-plane rotation through an angle $\gamma$. In practice, there is no preferred in-plane angle, so we assume that $\mu(R_z(\gamma)R)=\mu(R)$ for all $\gamma$, i.e., $\mu$ depends only on the viewing direction. We also augment the dataset with the reflection $I_j(x_1,-x_2)$ of every image, which is a projection from the antipodal viewing direction $-R_j^Te_3$, so $\mu$ takes the same value at antipodal viewing directions.

To keep $\mu$ nonnegative, we parametrize it as a normalized sum of squares
\begin{equation}
    \mu_B(R)=\frac{4\pi\sum_{i=1}^{K}p_B^{(i)}(R^Te_3)^2}{\sum_{i,l,m}|B_{i,l,m}|^2}
    \label{eqn:viewing_sos}
\end{equation}
of the spherical harmonic series
\begin{equation}
    p_B^{(i)}(\eta)=\sum_{l=0}^{P/2}\sum_{m=-l}^{l}B_{i,l,m}\,Y_l^m(\eta),\qquad i=1,\dots,K,
\end{equation}
where $K=(P/2+1)^2$, and $B_{i,l,-m}=\overline{B_{i,l,m}}$, so that each $p_B^{(i)}$ is real. Since the $Y_l^m$ are orthonormal on $\mathbb{S}^2$, $\int_{\SO(3)}\mu_B(R)\,dR=1$. To make $\mu_B$ symmetric under antipodal viewing directions, each $p_B^{(i)}$ contains only even or only odd degrees $l$. Since $\mu_B$ does not change when $B$ is scaled, we fix $B_{1,0,0}=1$.

We note that every nonnegative spherical harmonic series on $\mathbb{S}^2$ is a sum of squares of spherical harmonic series~\cite{scheiderer2006sums}. However, writing a nonnegative series of degree at most $P$ in this form may require some $p_B^{(i)}$ of degree higher than $P/2$. For simplicity, we restrict each $p_B^{(i)}$ to degree at most $P/2$, so $\mu_B$ in~\eqref{eqn:viewing_sos} may not represent every nonnegative density of degree at most $P$.

\subsubsection{Shift distribution representation}
\label{sec:shift_dist}
The offsets are written in polar coordinates as 
\begin{equation}\tau=(\rho\cos\psi,\rho\sin\psi).
\end{equation}
Since there is no preferred in-plane angle, we also assume that $\psi$ is uniform on $[0,2\pi)$. We assume that there exists $\rho_{\max}>0$ such  that $\rho\le\rho_{\max}$, and approximate the distribution of $\rho$ by a discrete distribution on the nodes \begin{equation}\rho_h=(h-1)\rho_{\max}/(N_\rho-1), \quad h=1,\dots,N_\rho.
\end{equation}

We represent the shift distribution by 
\begin{equation}
    \nu_\beta(\rho,\psi)=\sum_{h=1}^{N_\rho}\beta_h  \delta_{\rho_h,\rho}, \quad \psi \in [0,2\pi)
    \label{eqn:shift_density}
\end{equation}
where $\beta_h\ge0$ is the probability mass assigned to $\rho_h$ and $\sum_{h=1}^{N_\rho}\beta_h=1$.

\section{Method of moments of denoised images}
In this section, we present the details of MoMDeni. The method is based on image denoising using covariance Wiener filtering, which is described in Section~\ref{sec:fastpca}. In Section~\ref{sec:overcrowed}, we give a heuristic approach to detecting and discarding images containing more than one particle, to improve moment estimation. Finally, in Section~\ref{sec:kam}, we discuss reconstructing the volume from sample moments.

\subsection{Image denoising}
\label{sec:fastpca}

We briefly review the method of~\cite{marshall2023fast} which jointly denoises the images and corrects for the effects of their CTFs. It first estimates the covariance matrix of the clean projections from the observed noisy images~\cite{Zhao2014FastSP}, and then uses the estimated covariance to filter the images~\cite{bhamre2016denoising}.

Throughout, we group the coefficients with the same angular frequency $q$ into the vector $a_q=(a_{q,t})_{t=1}^{T_q}\in\C^{T_q}$, which corresponds to the Fourier mode $e^{iq\varphi}$. For each $q$, let $\tilde a_q^{(j)}$ be the coefficients of the observed image $I_j$ and $a_q^{(j)}$ be those of the clean image $\alpha_j \cP_{R_j} \Phi$. Assuming a radial CTF, we have 
\begin{equation}
   \tilde a_q^{(j)} = H_q^{(j)}  a_q^{(j)}+\epsilon_q^{(j)},
\end{equation}
where 
\begin{equation}
 (H_q^{(j)} )_{t,t'}=\int_0^c f_{q,t}(r) H_j(r) f_{q,t'}(r)\,r \,dr.   
\end{equation}
Since the images have already been whitened, we treat $\epsilon_q^{(j)}$ as white Gaussian noise with variance $\sigma^2=1$.

Since the rotations $R_j$ and contrasts $\alpha_j$ are random, the clean coefficients $a^{(j)}=(a^{(j)}_{q,t})_{q,t}$, $j=1,\dots,N$, are identically distributed samples of a random vector $a=(a_{q,t})_{q,t}$, where $N$ counts both the images and their reflections. We aim to estimate its mean $\mu=\E[a]$ and covariance
\begin{equation}
    C_{(q,t),(q',t')} := \E\big[(a_{q,t}-  \mu_{q,t}) \overline{(a_{q',t'}-  \mu_{q',t'})}\big].
\end{equation}
Since the in-plane rotations are uniform,  it can be shown that  $C_{(q,t),(q',t')}=0$ for $q\ne q'$~\cite{Zhao2014FastSP}. We therefore only need $\mu_0$ and the diagonal blocks of $C$,
\begin{equation}
    C_q := \E \big[(a_q-\mu_q) (a_q-\mu_q)^*\big]
    \label{eqn:cov}
\end{equation}
where the reflected images make each $C_q$ real. For each $q$, we estimate $\mu_q$ and $C_q$ from $\{\widetilde a_q^{(j)}\}_{j=1}^N$ by least squares~\cite{bhamre2016denoising}
\begin{equation}
\begin{split}
    \widehat\mu _q&=\operatorname*{arg\,min}_{\mu_q} \sum_{j=1}^N\| \widetilde a_q^{(j)} - H_q^{(j)} \mu _q \|_F^2,\\
    \widehat C_q&=\operatorname*{arg\,min}_{C_q} \sum_{j=1}^N\|(\tilde a_q^{(j)}-H_q^{(j)}\widehat \mu_q)(\tilde a_q^{(j)}-H_q^{(j)}\widehat \mu_q)^T
     -(H_q^{(j)}C_q(H_q^{(j)})^T+\sigma^2I_{T_q \times T_q} )\|_F^2,
\end{split}
\end{equation}
followed by eigenvalue shrinkage and the contrast-aware refinement of~\cite{shi2022contrast}.

With these estimates, the denoised coefficients are obtained by covariance Wiener filtering~\cite{bhamre2016denoising}
\begin{equation}
    \widehat a_q^{(j)}=  \widehat \mu_q+ \widehat C_q H_q^{(j)}\big(H_q^{(j)} \widehat C_q H_q^{(j)}+\sigma^2I_{T_q \times T_q}\big)^{-1} \big(\tilde a_q^{(j)}-H_q^{(j)} \widehat \mu_q\big).
\end{equation}

For efficient computation, we take $f_{q,t}$ to be the Fourier--Bessel basis, i.e., the Fourier transforms of the functions $J_{|q|}(z_{|q|,t} \rho) e^{iq\varphi}$ on a disk, where $J_{|q|}$ is the Bessel function of the first kind of order $|q|$ and $z_{|q|,t}$ is its $t$-th positive zero. For each $q$, the truncation limit $T_q$ is determined by the rule in~\cite{zhao2013fourier}. Using this basis, $H_q^{(j)}$ is diagonal and we can employ the fast Fourier-Bessel expansion algorithm of~\cite{marshall2023fle}.

\subsection{Discarding overcrowded images}
\label{sec:overcrowed}
\begin{figure}[ht!]
\centering
\includegraphics[width=0.5\linewidth]{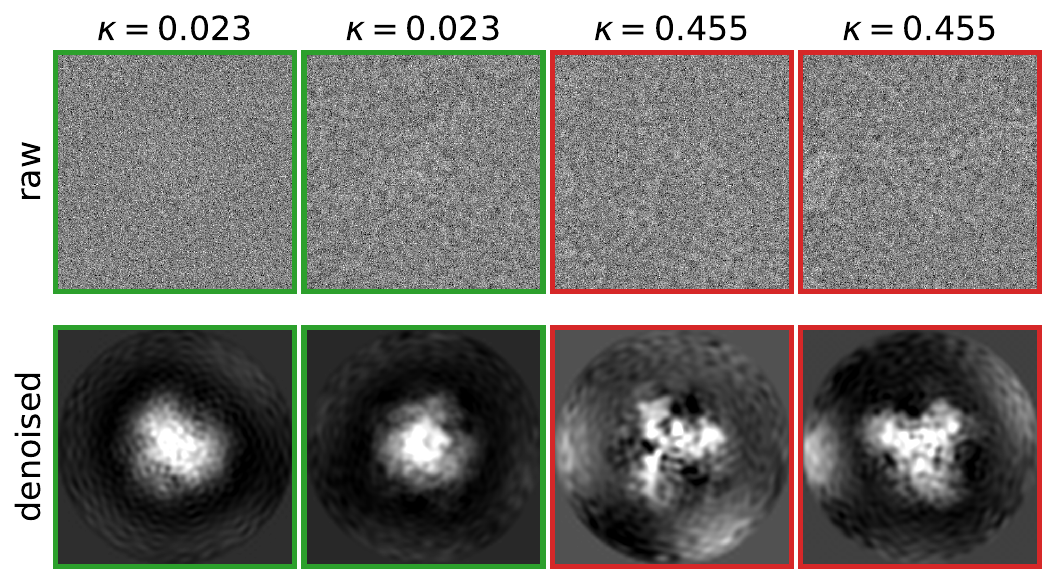}
\caption{Raw (top) and denoised (bottom) images from EMPIAR-10073~\cite{nguyen2016trisnrnp} with small (green) and large (red) crowding ratios $\kappa$ defined in~\eqref{eqn:crowding_factor}.  }
\label{fig:crowding}
\end{figure}

Our image model~\eqref{eqn:image_model} assumes that each image contains a single particle. However, particle pickers often produce images that also contain neighboring particles, and such images bias the moment estimates~\cite{andy2026overcrowded}. We therefore detect and discard these overcrowded images. The neighboring effect becomes more apparent in the
denoised images, as illustrated in Figure~\ref{fig:crowding}.

First, we use the coordinates of picked particles for initial screening. For each particle, we compute the distance to the nearest other particle picked on the same micrograph, and discard the particle if this distance is smaller than a fixed fraction of the box size. However, because of the high noise level, many neighboring particles can be missed by the particle picker, so a considerable fraction of the remaining images are still overcrowded.

We therefore use a heuristic criterion on the denoised images. For each smoothed denoised image $\hat I_j$, we consider two regions of the image: a central disk $\Omega_{\mathrm{in}}$, which contains the particle, and an outer annulus $\Omega_{\mathrm{out}}$ near the image boundary, where a neighboring particle would appear. For a region $\Omega$, let $S_\Omega\subset\Omega$ be the fraction $p_\Omega$ of its pixels with the largest values of $\hat I_j$ ($p_\Omega=20\%$ for the central disk and $10\%$ for the annulus). We measure the signal strength in $\Omega$ by their root mean square,
\begin{equation}
    E_\Omega(\hat I_j)=\Big(\frac{1}{|S_\Omega|}\sum_{x\in S_\Omega}\hat I_j(x)^2\Big)^{1/2}.
\end{equation}
Unlike the variance over the whole region, this measure focuses on the brightest pixels, so it is sensitive to a neighboring particle that occupies only part of the annulus. We define the crowding ratio
\begin{equation}
    \kappa_j=\frac{E_{\Omega_{\mathrm{out}}}(\hat I_j)}{E_{\Omega_{\mathrm{in}}}(\hat I_j)}.
    \label{eqn:crowding_factor}
\end{equation}
A large $\kappa_j$ indicates strong signal near the boundary relative to that in the central region. We keep the images with small $\kappa_j$, while keeping the proportions of images from different micrographs approximately the same as in the full dataset. We then repeat the denoising procedure in Section~\ref{sec:fastpca} using only the selected images to get higher-quality denoised images.

\subsection{Method of moments}
\label{sec:kam}

In this section, we describe reconstructing the volume from the moments of the denoised images. In Section~\ref{sec:moment}, we express the first three moments of clean projections in terms of the volume, the orientation distribution and the shift distribution. In Section~\ref{sec:moment_matching}, we compute sample moments from the denoised images and estimate these unknowns using numerical optimization.

\subsubsection{Moments from expansion coefficients}
\label{sec:moment}
To reconstruct the volume from the moments, we first show the mapping from unknown parameters $(A,B,\nu)$ to moments of clean projections. We start with the expansion coefficients of images $a_{q,t}(A,R,\tau)$ in~\eqref{eqn:proj_basis} with $\tau=(\rho\cos\psi,\rho\sin\psi)$. By~\eqref{eqn:shift}, the Fourier transform of the shifted projection is
\begin{equation}
    \mathcal{F}[\cT_\tau \cP_R\Phi_A ](r,\varphi)=e^{-2\pi i  r\rho \cos(\varphi-\psi)} \mathcal{F}[ \cP_R\Phi_A ](r,\varphi).
    \label{eqn:shift_phase}
\end{equation}
The Jacobi--Anger expansion~\cite{abramowitz1964handbook} gives the following expansion of the phase factor
\begin{equation}
    e^{-2\pi i  r\rho \cos(\varphi-\psi)}=\sum_{u=-\infty}^{\infty}(-i)^u J_u(2\pi r\rho)\,e^{iu(\varphi-\psi)},
    \label{eqn:jacobi_anger}
\end{equation}
where $J_u$ is the Bessel function of the first kind. Since $a_{q',t'}(A,R,0)=0$ for $|q'|>L$, the unshifted projection is
\begin{equation}
    \mathcal F[\cP_R\Phi_A](r,\varphi)
    =\sum_{q'=-L}^{L}\sum_{t'=1}^{T_{q'}}a_{q',t'}(A,R,0)\,f_{q',t'}(r)\,e^{iq'\varphi}.
\end{equation}
Substituting this and~\eqref{eqn:jacobi_anger} into~\eqref{eqn:shift_phase} gives
\begin{equation}
    \begin{aligned}
    &\mathcal F[\cT_\tau\cP_R\Phi_A](r,\varphi)=\sum_{u=-\infty}^{\infty}\sum_{q'=-L}^{L}\sum_{t'=1}^{T_{q'}}(-i)^u e^{-iu\psi}J_u(2\pi r\rho) a_{q',t'}(A,R,0)\,f_{q',t'}(r)\,e^{i(q'+u)\varphi}.
    \end{aligned}
\end{equation}
The coefficients of the shifted projection are given by
\begin{equation}
    \begin{aligned}
    a_{q,t}(A,R,\tau)&=\frac{1}{2\pi}\int_0^{2\pi}\int_0^c\mathcal F[\cT_\tau\cP_R\Phi_A](r,\varphi)f_{q,t}(r)\,e^{-iq\varphi}\,r\,dr\,d\varphi\\
    &=\sum_{q'=-L}^{L}\sum_{t'=1}^{T_{q'}}(-i)^{q-q'}e^{-i(q-q')\psi}\big(G_{q,q'}(\rho)\big)_{t,t'}\,a_{q',t'}(A,R,0),
    \label{eqn:coef_shift}
    \end{aligned}
\end{equation}
where the matrices $G_{q,q'}(\rho)\in\R^{T_q\times T_{q'}}$ have the entries
\begin{equation}
    \big(G_{q,q'}(\rho)\big)_{t,t'}= \int_0^c f_{q,t}(r)\,f_{q',t'}(r)  J_{q-q'}(2\pi r\rho)\,r\,dr.
\end{equation}

\begin{table}[ht!]
\centering
\setlength{\tabcolsep}{4pt}
\begin{tabular}{lccccc}
\hline
 & Sample size & Box size (pixels) & Pixel size (\AA) & Weight (kDa)& Reference map \\
\hline
EMPIAR-10017 & 40863 & 160 & 1.77 & 465 & \href{https://www.ebi.ac.uk/emdb/EMD-2824}{EMD-2824} \\
EMPIAR-10028 & 105247 & 360 & 1.34 &  3000 &\href{https://www.ebi.ac.uk/emdb/EMD-2660}{EMD-2660} \\
EMPIAR-10073 & 138899 & 380 & 1.43 & 1000 &\href{https://www.ebi.ac.uk/emdb/EMD-8012}{EMD-8012} \\
EMPIAR-10081 & 55870 & 256 & 1.30 & 298 &\href{https://www.ebi.ac.uk/emdb/EMD-8511}{EMD-8511} \\
EMPIAR-11005 & 57148 & 360 & 1.16 & 2000 & \href{https://www.ebi.ac.uk/emdb/EMD-14113}{EMD-14113}\\
EMPIAR-11618 & 68401 & 256 & 0.952 & 206 &\href{https://www.ebi.ac.uk/emdb/EMD-17944}{EMD-17944} \\
\hline
\end{tabular}
\caption{Information about the datasets used in Section~\ref{sec:numerics} including sample sizes, box sizes, pixel sizes, molecular weights and corresponding reference maps. We select $10^4$ denoised images from the raw images to estimate the sample moments in Section~\ref{sec:numerics}.}
\label{tab:datasets}
\end{table}

For a volume, an orientation distribution and a shift distribution given by $(A,B,\beta)$, we define the first three population moments of the steerable coefficients of the images as
\begin{align}
    M_1[A,B,\beta](q,t) &= \int_{\mathcal{SO}(3) \times \mathbb{R}^2}a_{q,t}(A,R,\tau)\mu_B(R) \nu_\beta(\tau)\,dR\,d\tau ,
    \label{eqn:M1_population}\\
    M_2[A,B,\beta](q_1,t_1;q_2,t_2)&=\int_{\mathcal{SO}(3) \times \mathbb{R}^2}a_{q_1,t_1}(A,R,\tau)\, \overline{a_{q_2,t_2}(A,R,\tau)}\mu_B(R) \nu_\beta(\tau)\,dR\,d\tau,
    \label{eqn:M2_population}\\
    M_3[A,B,\beta](q_1,t_1;q_2,t_2;q_3,t_3)&=\int_{\mathcal{SO}(3) \times \mathbb{R}^2}a_{q_1,t_1}(A,R,\tau)\, a_{q_2,t_2}(A,R,\tau)a_{q_3,t_3}(A,R,\tau) \nonumber\\ 
    & \qquad \times  \mu_B(R) \nu_\beta(\tau)\,dR\,d\tau
    \label{eqn:M3_population}
\end{align}
where $\mu_B$ and $\nu_\beta$ are given in~\eqref{eqn:viewing_sos} and~\eqref{eqn:shift_density}.

\begin{figure}[ht!]
\centering
\begin{subfigure}{0.25\textwidth}
    \centering
    \includegraphics[width=\linewidth]{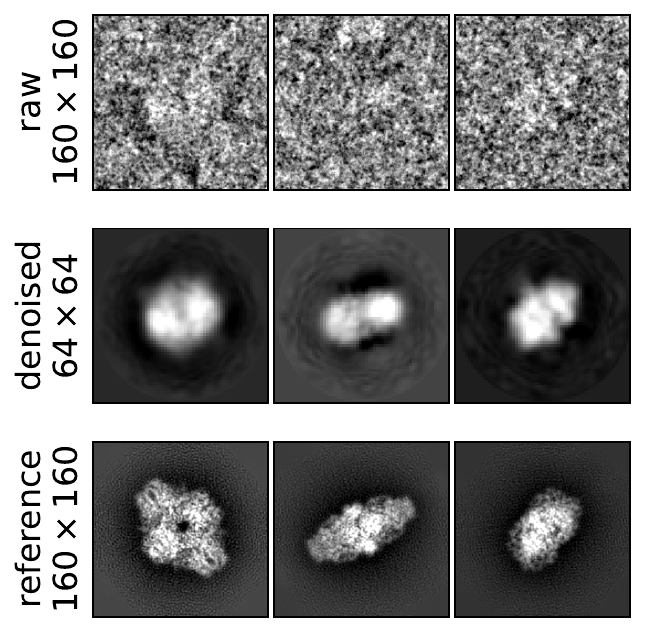}
    \caption{EMPIAR-10017}
\end{subfigure}
\begin{subfigure}{0.25\textwidth}
    \centering
    \includegraphics[width=\linewidth]{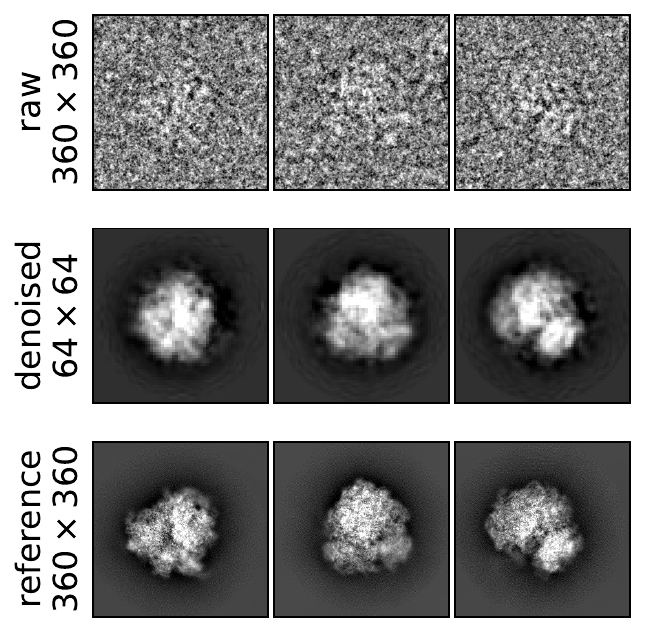}
    \caption{EMPIAR-10028}
\end{subfigure}
\begin{subfigure}{0.25\textwidth}
    \centering
    \includegraphics[width=\linewidth]{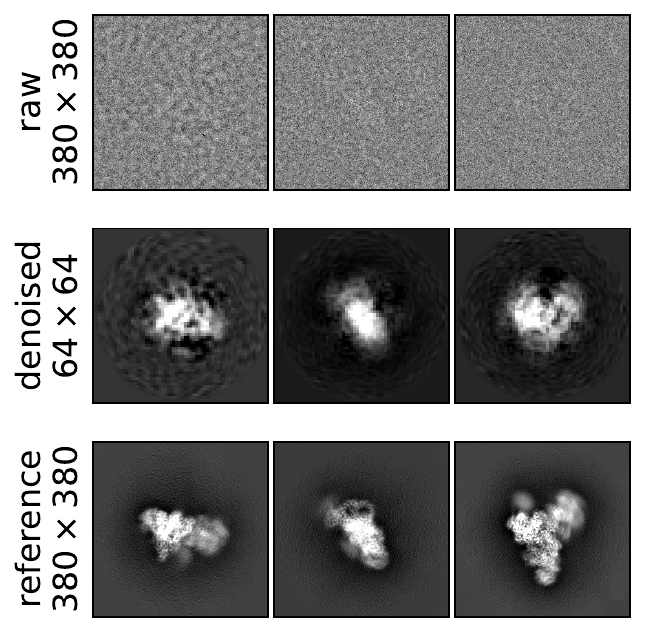}
    \caption{EMPIAR-10073}
\end{subfigure}

\begin{subfigure}{0.25\textwidth}
    \centering
    \includegraphics[width=\linewidth]{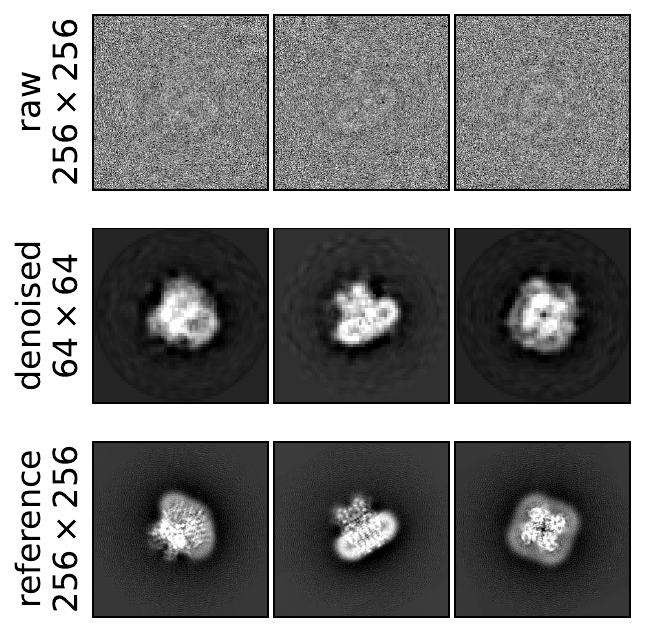}
    \caption{EMPIAR-10081}
\end{subfigure}
\begin{subfigure}{0.25\textwidth}
    \centering
    \includegraphics[width=\linewidth]{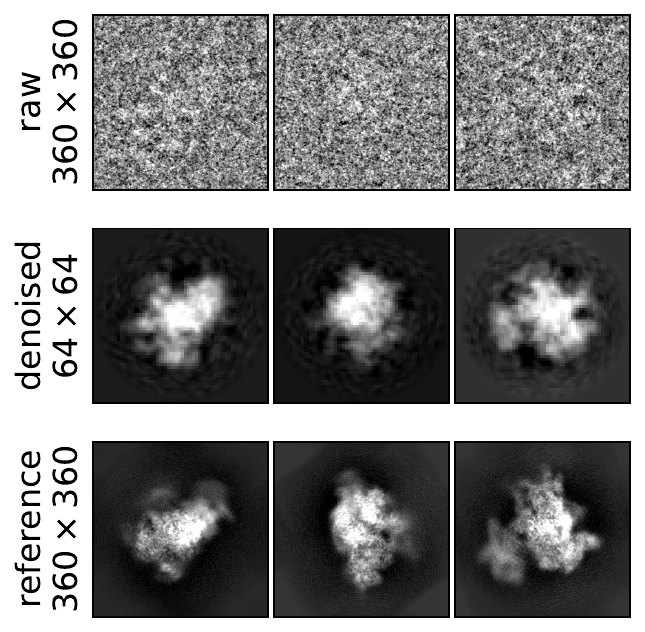}
    \caption{EMPIAR-11005}
\end{subfigure}
\begin{subfigure}{0.25\textwidth}
    \centering
    \includegraphics[width=\linewidth]{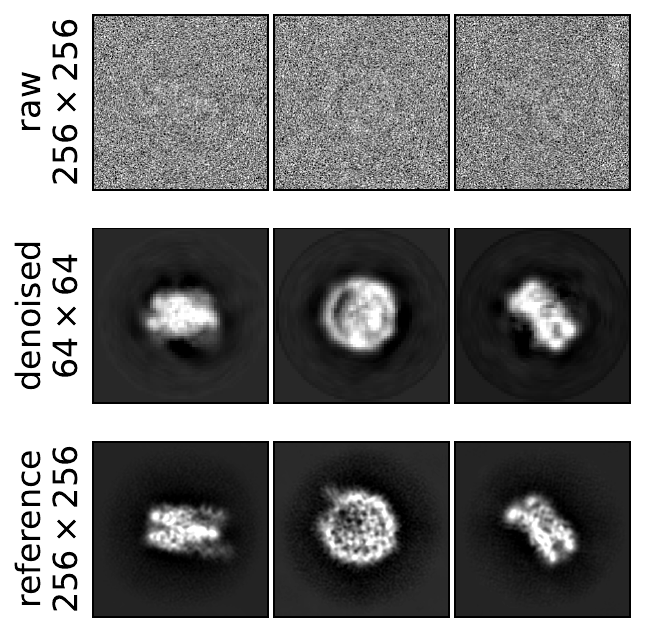}
    \caption{EMPIAR-11618}
\end{subfigure}

\caption{Raw particle images (top rows), the corresponding denoised images used to estimate the moments (middle rows), and clean projections of the corresponding reference maps (bottom rows).}
\label{fig:denoised_examples}
\end{figure}

Because the image distribution is invariant under in-plane rotations, most entries of the moment tensors vanish. To see this, we use the fact that rotating an image in-plane by $\gamma$ rotates both the projection and the shift. Let 
\begin{equation}\tau_\gamma=(\rho\cos(\psi+\gamma),\rho\sin(\psi+\gamma))
\end{equation}
be the shift rotated by $\gamma$. In~\eqref{eqn:coef_shift}, replacing $\psi$ by $\psi+\gamma$ gives the factor $e^{-i(q-q')\gamma}$, and replacing $R$ by $R_z(\gamma)R$ gives the factor $e^{-iq'\gamma}$, so
\begin{equation}
a_q(A,R_z(\gamma)R,\tau_\gamma)=e^{-iq\gamma}\,a_q(A,R,\tau).
\end{equation}
Since $\mu_B$ is invariant under in-plane rotations and $\psi$ is uniform and independent of $R$, $(R_z(\gamma)R,\tau_\gamma)$ has the same distribution as $(R,\tau)$. Hence, for every $\gamma$,
\begin{equation}
    \begin{aligned}
    M_3[A,B,\beta](q_1,t_1;q_2,t_2;q_3,t_3)&=\int_{\mathcal{SO}(3)\times \mathbb{R}^2}\Big[\prod_{i=1}^{3}a_{q_i,t_i}(A,R_z(\gamma)R,\tau_\gamma)\Big]\mu_B(R)\nu_\beta(\tau)\,dR\,d\tau\\
    &=e^{-i(q_1+q_2+q_3)\gamma}\,M_3[A,B,\beta](q_1,t_1;q_2,t_2;q_3,t_3),
    \end{aligned}
\end{equation}
so $M_3$ has to vanish unless $q_1+q_2+q_3=0$. In the same way, we can show that $M_1$ vanishes unless $q=0$ and $M_2$ vanishes unless $q_1=q_2$. Moreover, the integral over the in-plane angle can be computed analytically, and only the viewing directions need to be discretized. With weights $w_k$ and rotations $\widetilde R_k$, $k=1,\dots,K$, where $\widetilde R_k$ maps the $z$-axis to the $k$-th node of a quadrature rule on $\mathbb{S}^2$~\cite{graf2009sampling,graf2013efficient}, the moments become
\begin{align}
    &M_1[A,B,\beta](q,t)=\delta_{q,0}\sum_{h=1}^{N_\rho}\beta_h\sum_{k=1}^{K}w_k\,\mu_B(\widetilde R_k)\big[G_{0,0}(\rho_h)\,a_0(A,\widetilde R_k,0)\big]_t,\label{eqn:m1quad}\\
    &M_2[A,B,\beta](q_1,t_1;q_2,t_2)=\delta_{q_1,q_2}\sum_{h=1}^{N_\rho}\beta_h\sum_{k=1}^{K}w_k\,\mu_B(\widetilde R_k)\sum_{q'=-L}^{L}\big[G_{q_1,q'}(\rho_h)\,a_{q'}(A,\widetilde R_k,0)\big]_{t_1}\nonumber\\*
    &\qquad\times\overline{\big[G_{q_1,q'}(\rho_h)\,a_{q'}(A,\widetilde R_k,0)\big]_{t_2}},\label{eqn:m2quad}\\
    &M_3[A,B,\beta](q_1,t_1;q_2,t_2;q_3,t_3)=\delta_{q_1+q_2+q_3,0}\sum_{h=1}^{N_\rho}\beta_h\sum_{k=1}^{K}w_k\,\mu_B(\widetilde R_k) \nonumber \\ 
    &\qquad \times \sum_{q_1'+q_2'+q_3'=0}\ \prod_{i=1}^{3}\big[G_{q_i,q_i'}(\rho_h)\,a_{q_i'}(A,\widetilde R_k,0)\big]_{t_i},\label{eqn:m3quad}
\end{align}
where $[\,\cdot\,]_t$ denotes the $t$-th entry of a vector. We provide the detailed derivation in Appendix~\ref{app:moment_derivation}.

\subsubsection{Reconstruction from sample moments}
\label{sec:moment_matching}
Let $\hat a^{(j)}_{q,t}$, $j=1,\dots,N$, be the coefficients of the denoised and selected images. We form the first three sample moments
\begin{align}
\widehat M_1(q,t)&=\frac{\delta_{q,0}}{N}\sum_{j=1}^{N}\widehat a^{(j)}_{q,t},\\
    \widehat M_2(q_1,t_1;q_2,t_2)&=\frac{\delta_{q_1,q_2}}{N}\sum_{j=1}^{N}\widehat a^{(j)}_{q_1,t_1}\,\overline{\widehat a^{(j)}_{q_2,t_2}},\\
    \widehat M_3(q_1,t_1;q_2,t_2;q_3,t_3)&=\frac{\delta_{q_1+q_2+q_3,0}}{N}\sum_{j=1}^{N}\widehat a^{(j)}_{q_1,t_1}\,\widehat a^{(j)}_{q_2,t_2}\,\widehat a^{(j)}_{q_3,t_3}.
\end{align}

\begin{figure}[ht!]
\centering
\includegraphics[width=0.75\textwidth]{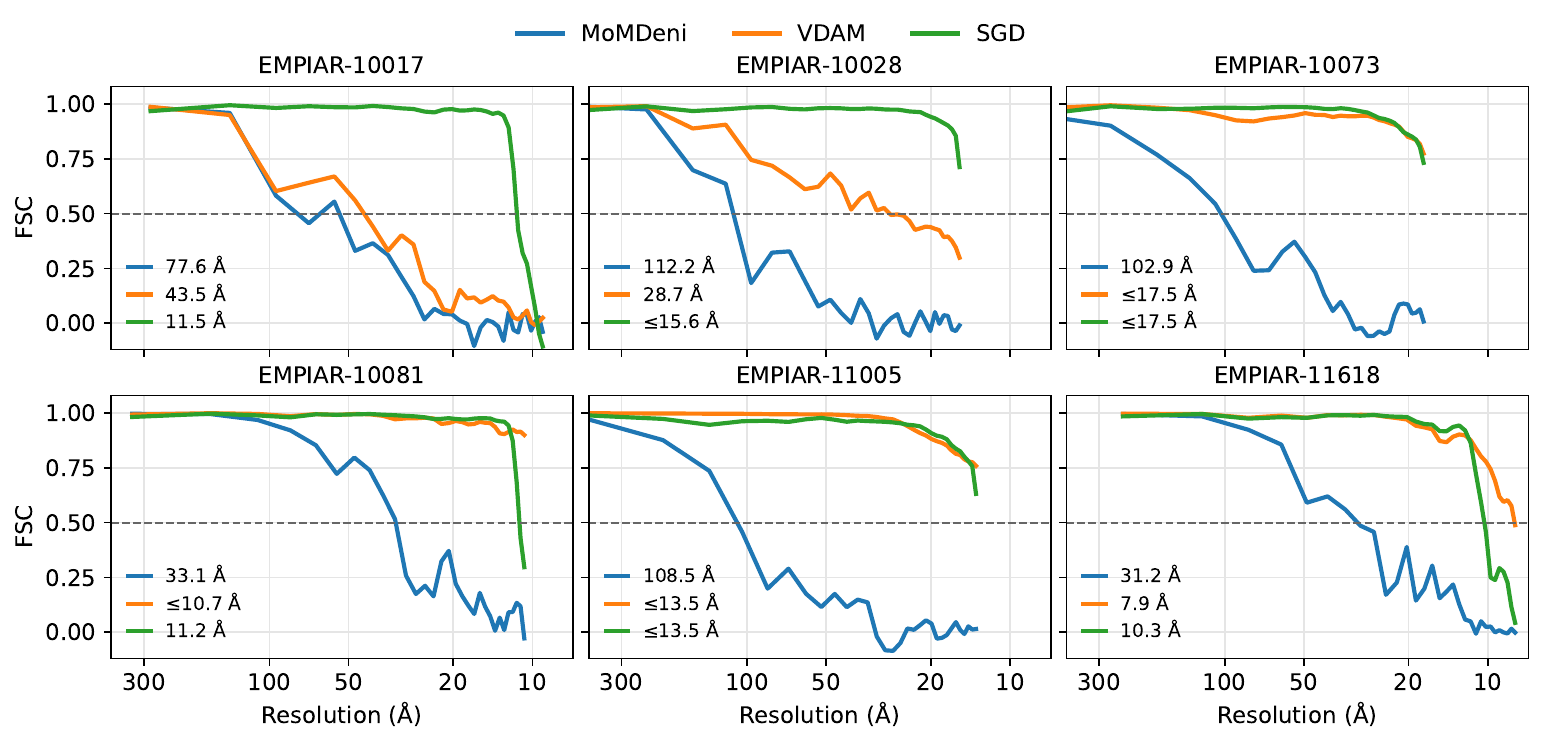}
\caption{Fourier-shell correlation (FSC) curves between the \textit{ab-initio} models obtained by MoMDeni (our method), VDAM (RELION) and SGD (cryoSPARC) and the corresponding reference maps. Resolutions are determined using an FSC threshold of $1/2$.}
\label{fig:fsc_abinitio}
\end{figure}

We define the loss on the $k$-th moment as
\begin{equation}
    \ell_k(A,B,\beta)=\frac{\|M_k[A,B,\beta]-\hat M_k\|_F^2}{\|\hat M_k\|_F^2}.
\end{equation}
Following~\cite{doi:10.1137/24M1699644,katsevich2023likelihood,Fan2024Maximum}, we match the moments sequentially, minimizing $\ell_1$, $\ell_2$ and $\ell_3$ in turn, each stage initialized by the solution of the previous one.  We solve each stage with the L-BFGS-B method~\cite{byrd1995limited}. In particular, we use 10 initializations for the sequential minimization of $\ell_1$ and $\ell_2$. For each initialization, we run until the convergence criterion is met. We then select the parameter set that yields the lowest value of $\ell_3$ and use it to initialize the minimization of $\ell_3$, which we run for at most 500 iterations.

\begin{table}[ht!]
\centering
\small
\begin{tabular}{lccc}
\hline
EMPIAR & from MoMDeni & from VDAM & from SGD \\
\hline
10017 & \shortstack{$\mathbf{16.1}$ (3.6 + 12.5) mins\\$\mathbf{21}$ iters} & \shortstack{$\mathbf{15.1}$ (5.4 + 9.7) mins\\$\mathbf{14}$ iters} & \shortstack{$\mathbf{14.0}$ (3.1 + 10.9) mins\\$\mathbf{18}$ iters} \\[3pt]
10028 & \shortstack{$\mathbf{108.2}$ (5.3 + 102.9) mins\\$\mathbf{27}$ iters} & \shortstack{$\mathbf{107.9}$ (17.1 + 90.8) mins\\$\mathbf{23}$ iters} & \shortstack{$\mathbf{82.2}$ (3.8 + 78.4) mins\\$\mathbf{18}$ iters} \\[3pt]
10073 & \shortstack{$\mathbf{121.8}$ (8.8 + 113.0) mins\\$\mathbf{28}$ iters} & \shortstack{$\mathbf{105.9}$ (24.5 + 81.4) mins\\$\mathbf{18}$ iters} & \shortstack{$\mathbf{90.6}$ (3.8 + 86.8) mins\\$\mathbf{18}$ iters} \\[3pt]
10081 & \shortstack{$\mathbf{22.6}$ (6.2 + 16.4) mins\\$\mathbf{15}$ iters} & \shortstack{$\mathbf{24.2}$ (6.5 + 17.7) mins\\$\mathbf{18}$ iters} & \shortstack{$\mathbf{20.1}$ (3.5 + 16.6) mins\\$\mathbf{15}$ iters} \\[3pt]
11005 & \shortstack{$\mathbf{44.4}$ (6.4 + 38.0) mins\\$\mathbf{26}$ iters} & \shortstack{$\mathbf{38.8}$ (7.2 + 31.6) mins\\$\mathbf{19}$ iters} & \shortstack{$\mathbf{24.8}$ (3.8 + 21.0) mins\\$\mathbf{15}$ iters} \\[3pt]
11618 & \shortstack{$\mathbf{43.1}$ (8.5 + 34.6) mins\\$\mathbf{20}$ iters} & \shortstack{$\mathbf{43.1}$ (10.9 + 32.2) mins\\$\mathbf{19}$ iters} & \shortstack{$\mathbf{32.1}$ (2.7 + 29.4) mins\\$\mathbf{17}$ iters} \\[3pt]
\hline
\end{tabular}
\caption{Total runtime (\textit{ab-initio} modeling + RELION refinement) and number of refinement iterations on the experimental datasets. }
\label{tab:runtime}
\end{table}

\section{Numerical results}
\label{sec:numerics}

In this section, we show the first successful application of the method-of-moments-based \textit{ab-initio} modeling without assuming a uniform distribution of viewing directions, to six publicly available experimental datasets of different macromolecular complexes: 
\begin{itemize}
    \item EMPIAR-10017~\cite{SCHERES2015114}: $\beta$-galactosidase.
    \item EMPIAR-10028~\cite{10.7554/eLife.03080}: 80S ribosome.
    \item EMPIAR-10073~\cite{nguyen2016trisnrnp}: Yeast U4/U6.U5 tri-snRNP.
    \item EMPIAR-10081~\cite{lee2017hcn1}: HCN1 ion channel. 
    \item EMPIAR-11005~\cite{10.1093/nar/gkac283}: Human 48S translation initiation complex.
    \item EMPIAR-11618~\cite{bacic2021alc1}: Nucleosome.
\end{itemize}
 Table~\ref{tab:datasets} lists the number of images, box size, pixel size, molecular weight, and reference map of each dataset. We use the particle picks deposited in EMPIAR, except for EMPIAR-10017, for which we use the particle picks provided in~\cite{Dhakal2023}. EMPIAR-11005 contains two subsets: one corresponding to human 48S IC in the open state and the other corresponding to human 48S IC in the closed state. We use the first subset.

\begin{figure}[ht!]
\centering
\includegraphics[width=0.75\textwidth]{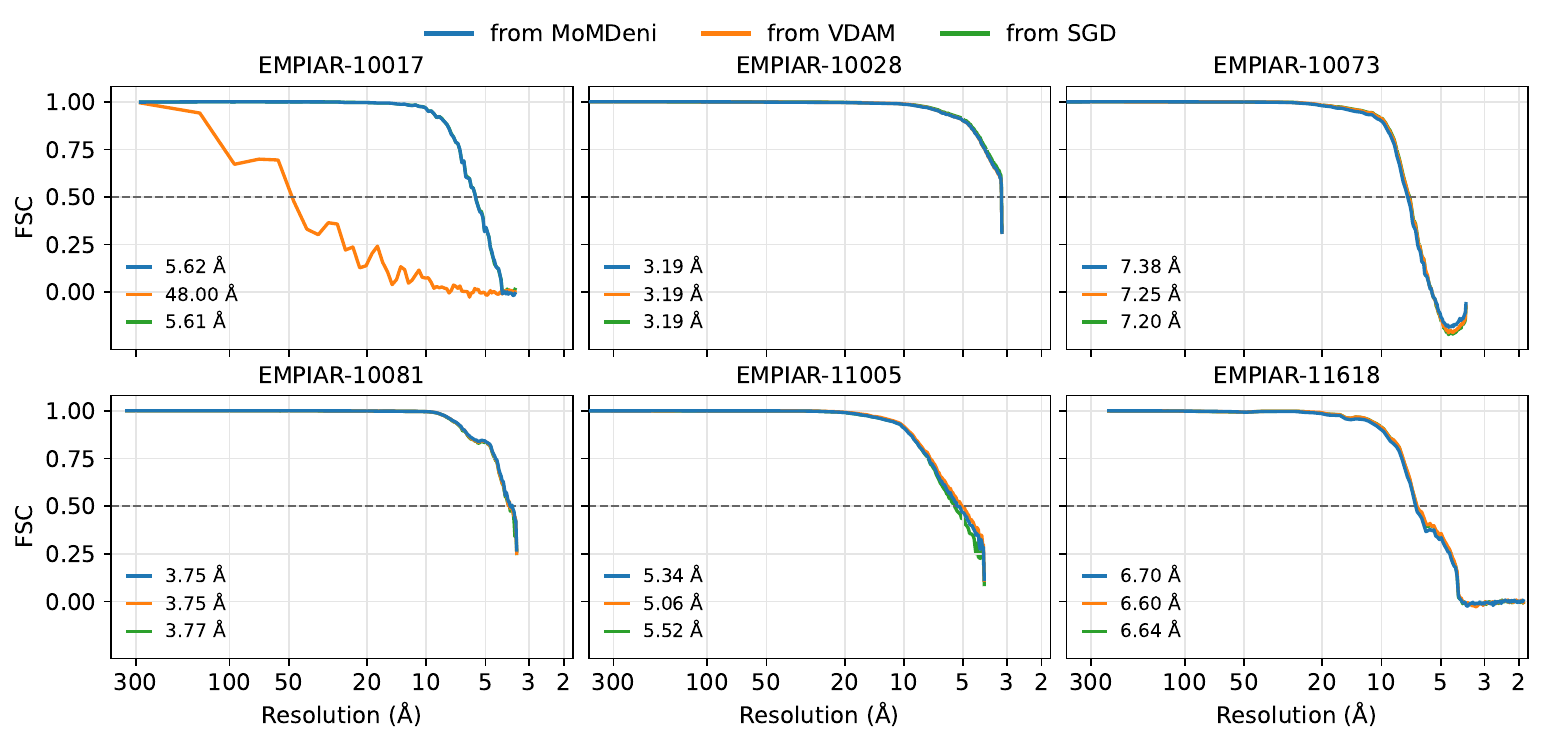}
\caption{FSC curves between the refined models initialized by MoMDeni (our method), VDAM (RELION) and SGD (cryoSPARC) and the corresponding reference maps. Resolutions are determined using an FSC threshold of $1/2$.}
\label{fig:fsc_refine}
\end{figure}

We impose $\mathbf{D}_2$ symmetry for EMPIAR-10017 and $\mathbf{C}_4$  symmetry for EMPIAR-10081 during both \textit{ab-initio} modeling and refinement. We downsample the images to $64\times 64$ pixels for \textit{ab-initio} modeling (Section~\ref{sec:abinitio}) and perform refinement (Section~\ref{sec:refinement}) on original particle images at their original resolution.  All \textit{ab-initio} experiments are run on a single NVIDIA H100 GPU (80~GB), and the refinements use two GPUs of the same type.

 \subsection{Results of \textit{ab-initio} modeling}
 \label{sec:abinitio}
\begin{figure}[ht!]
\centering
\includegraphics[width=0.75\textwidth]{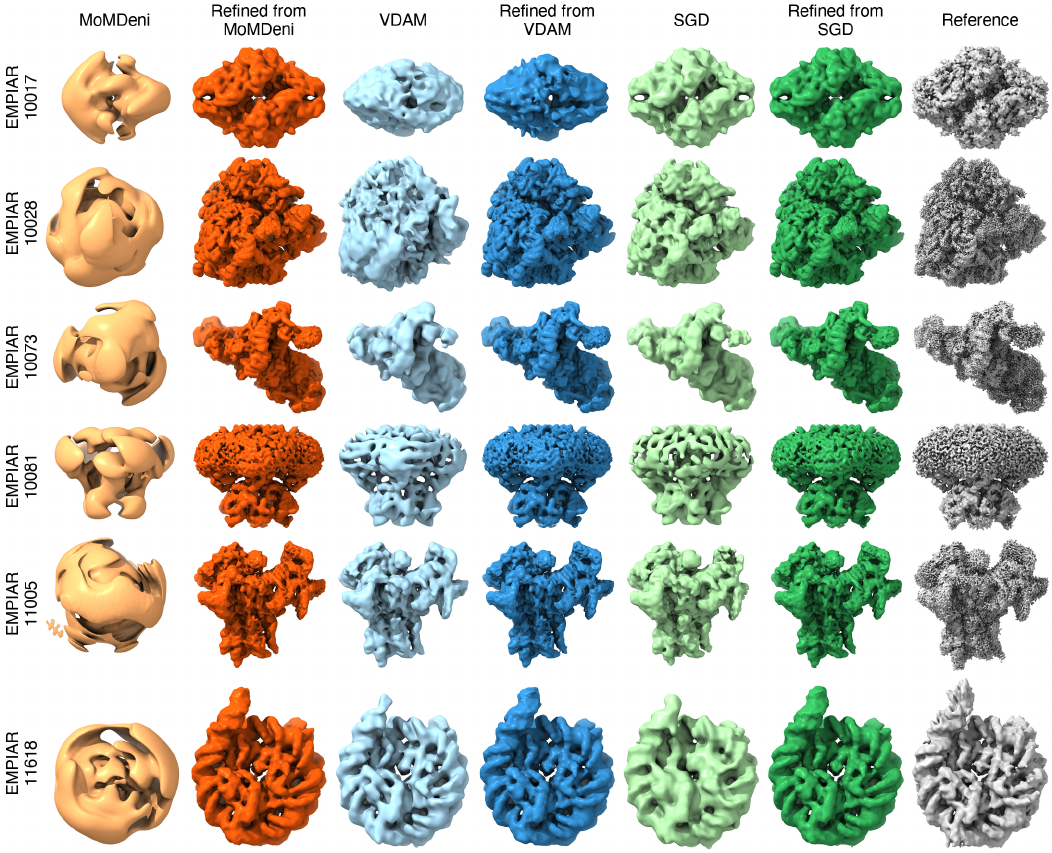}
\caption{Visualization of the \textit{ab-initio} models (our MoMDeni, VDAM and SGD),  the refined models initialized by the \textit{ab-initio} models and the reference maps rendered by UCSF ChimeraX~\cite{https://doi.org/10.1002/pro.4792}.}
\label{fig:volumes}
\end{figure}

 For each dataset, we select $10^4$ denoised images to estimate the sample moments using the method in Section~\ref{sec:overcrowed}. In Figure~\ref{fig:denoised_examples}, we show examples of denoised images and the corresponding raw images. The denoised images have low resolution but capture the main structural feature of the molecules (see Figure~\ref{fig:volumes}). 

 In our reconstruction, we truncate the expansion of  $\Phi_A$ in~\eqref{eqn:sfb3d} at $L=7$ and choose the radial functions $F_{l,s}$ to be the rescaled spherical Bessel functions used in~\cite{bendory2023autocorrelation}.  We set $P=2$ in the representation of the viewing direction density $\mu_B$ in~\eqref{eqn:viewing_sos}. For the shift distribution  $\nu_\beta$ in~\eqref{eqn:shift_density}, we set $\rho_{\max}$ to be $1/8$ of the box size (8 pixels), and assign probability masses to nodes spaced half a pixel apart. As a result, there are $467$ coefficients for $\Phi_A$, $6$ coefficients for $\mu_B$, and $17$ coefficients for $\nu_\beta$.

For comparison, we run two \textit{ab-initio} methods based on stochastic optimization: the variable-metric gradient descent with adaptive moments estimation (VDAM)~\cite{10.1042/BCJ20210708} in RELION 5.0.1 and stochastic gradient descent (SGD)~\cite{punjani2017cryosparc} in cryoSPARC 5.0.6, respectively, using the default settings of each package. Because of limitations in image denoising, the models reconstructed by MoMDeni have lower resolution than those obtained by VDAM or SGD, as shown in Figure~\ref{fig:fsc_abinitio}. In Table~\ref{tab:runtime},  we report the running time of the three methods. MoMDeni is faster than VDAM but slower than the highly optimized SGD method in cryoSPARC.

\subsection{Results of refinement} 
\label{sec:refinement}
We then use these \textit{ab-initio} models to initialize high-resolution refinement in RELION 5.0.1 using its default settings. Also in Table~\ref{tab:runtime}, we report total runtime and number of refinement iterations. Refinement initialized with the lower-resolution MoMDeni models generally requires a few more iterations but the total runtime remains comparable to that of the other two methods. We note that RELION applies a 60 Å low-pass filter to all initial models by default. This suppresses finer details in the higher-resolution model and may partly explain the comparable runtimes despite the lower initial resolutions of the MoMDeni models.

In Figure~\ref{fig:fsc_refine}, we show that refinements initialized with the MoMDeni models converge to high-resolution maps of comparable quality to those obtained from SGD-initialized models. For EMPIAR-10017, refinement initialized with VDAM fails to converge to the true structure. In Figure~\ref{fig:volumes}, we visualize the \textit{ab-initio} models, the refined models  and the reference maps.

These results demonstrate that, perhaps surprisingly, despite their lower initial resolutions, the MoMDeni models capture sufficient structural information to serve as effective initializations for high-resolution refinement. These results for method-of-moments-based \textit{ab-initio} modeling on experimental datasets are still preliminary. We discuss the limitations and directions for future work in Section~\ref{sec:discussion}.

\section{Discussion}
\label{sec:discussion}
For the first time, we demonstrated that the method of moments assuming a non-uniform distribution of orientations can produce \textit{ab-initio} models from experimental cryo-EM data that successfully initialize iterative refinement. Since the moments are computed in a single pass over the selected denoised images, without estimating the orientation of each image, our method produces these models rapidly. Several limitations remain to be addressed.

\begin{figure}[ht!]
\centering
\includegraphics[width=0.3\linewidth]{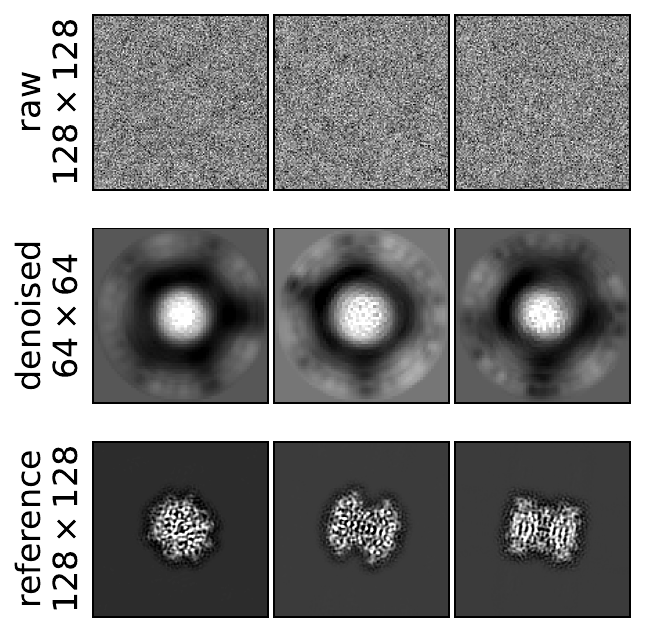}
\caption{Example of image denoising on EMPIAR-10335~\cite{doi:10.1073/pnas.1919114117}. The  weight of the molecule (\href{https://www.ebi.ac.uk/emdb/EMD-20907}{EMD-20907}) is only 52 kDa. The denoised images do not  reveal discernible structural features.}
\label{fig:denoise_fail}
\end{figure}

Our image denoising method is far from optimal. We use it because reliable estimation of the moments from raw experimental images remains challenging, particularly for the third moment. Compared with 2-D class averaging, covariance Wiener filtering suppresses noise at a low computational cost, and the denoised images help identify overcrowded images. However, denoising changes the image statistics, and the resulting moments are biased estimates of the moments of the clean projections. The quality of the denoised images is also limited. As shown in Figure~\ref{fig:denoised_examples}, the current denoising method preserves only the main structural features and loses finer details. If too little structural information remains, the subsequent moment-based reconstruction fails. Figure~\ref{fig:denoise_fail} illustrates this limitation for a target of low molecular weight. Improving image denoising may extend the range of datasets that our method can process.

An important future direction is to estimate the moments directly from the raw images. For the second moment, improved covariance estimation and eigenvalue shrinkage provide useful tools~\cite{bhamre2016denoising}. For the third moment, similar tools remain to be developed. More generally,  regularizing the moments~\cite{doi:10.1073/pnas.2419140122}, rather than denoising the images, may be more effective. Together, these would remove the need for imperfect image denoising.

Several factors also contribute to the mismatch between experimental data and our forward model. First, the image scale and background mean vary considerably across images and cannot be removed by normalizing each image~\cite{shi2022contrast,scheres2009maximum}. Second, our noise model is not entirely accurate. It assumes that the noise is added after the CTF and has the same power spectrum in all images. However, the images also contain structural noise from the ice (and the carbon support, if present), which is itself modulated by the CTF~\cite{zeng2007maximum}. Thon rings appear even in images of pure ice~\cite{mcmullan2015thon}. Hence, the noise spectrum varies with the defocus of each micrograph, whereas we whiten all images with a single spectrum (Section~\ref{sec:image_model}). Since some micrographs contain only a few picked particles, the spectrum of each micrograph should be estimated from the image margins together with additional particle-free regions of the micrograph. Third, the particles in publicly available datasets are often polished~\cite{scheres2014beam,zivanov2019bayesian}. Each particle is motion-corrected individually, and its movie frames are summed with frequency-dependent weights that are not part of the CTF. Afterwards, each extracted particle image is normalized so that its background has zero mean and unit variance. These steps are designed for iterative refinement, and we conjecture that some of them may not be suitable for the method of moments.

Future work should therefore close this gap from both sides. On the modeling side, these factors can be incorporated into the method of moments. For example, the image scales and the background means can be modeled as random variables and estimated jointly with the volume, the orientation distribution and the shift distribution. The overcrowding can be corrected in a more principled way~\cite{andy2026overcrowded}. On the data side, we plan to develop a method-of-moments pipeline for particles extracted from motion-corrected micrographs~\cite{li2013electron,zheng2017motioncor2}, without further polishing at the micrograph level or normalization at the particle level. The moments could also be estimated directly from the micrographs without particle picking~\cite{bendory2023toward}, which is particularly appealing for small molecules that are hard to pick.

Future work should also consider the total computational cost of \textit{ab-initio} modeling and the subsequent refinement. As moment-based \textit{ab-initio} models reach higher resolution, the initial low-pass filter in RELION could use a higher-resolution cutoff, determined from the resolution of these models, to shorten the refinement. Our GPU implementation could also be accelerated further, for example with CUDA.

\section*{Acknowledgments}
Y.S. was supported by NSF DMS 2514152. A.S., Y.W. and A.Z. were supported in part by NSF DMS 2510039, CZ Biohub FP00007575, and the Simons Foundation Math+X Investigator Award. The authors thank Eitan Levin for helpful discussion on the use of denoised images.

Generative AI (Claude Opus 5.5) was used to help code implementation and language editing. The authors reviewed all outputs and take full responsibility for the work.

\bibliographystyle{unsrt}
\bibliography{references}

\begin{thebibliography}{10}

\bibitem{shen2018nobel}
Peter~S. Shen.
\newblock The 2017 {N}obel {P}rize in {C}hemistry: cryo-{EM} comes of age.
\newblock {\em Anal. Bioanal. Chem.}, 410:2053--2057, 2018.

\bibitem{Nakane2020}
Takanori Nakane, Abhay Kotecha, Andrija Sente, Greg McMullan, Simonas Masiulis, Patricia M. G.~E. Brown, Ioana~T. Grigoras, Lina Malinauskaite, Tomas Malinauskas, Jonas Miehling, Tomasz Uchanski, Lingbo Yu, Dimple Karia, Evgeniya~V. Pechnikova, Erwin de~Jong, Jeroen Keizer, Maarten Bischoff, Jamie McCormack, Peter Tiemeijer, Steven~W. Hardwick, Dimitri~Y. Chirgadze, Garib Murshudov, A.~Radu Aricescu, and Sjors H.~W. Scheres.
\newblock Single-particle cryo-{EM} at atomic resolution.
\newblock {\em Nature}, 587(7832):152--156, 2020.

\bibitem{Yip2020}
Ka~Man Yip, Niels Fischer, Elham Paknia, Ashwin Chari, and Holger Stark.
\newblock Atomic-resolution protein structure determination by cryo-{EM}.
\newblock {\em Nature}, 587(7832):157--161, 2020.

\bibitem{bepler2019positive}
Tristan Bepler, Andrew Morin, Micah Rapp, Julia Brasch, Lawrence Shapiro, Alex~J. Noble, and Bonnie Berger.
\newblock Positive-unlabeled convolutional neural networks for particle picking in cryo-electron micrographs.
\newblock {\em Nat. Methods}, 16:1153--1160, 2019.

\bibitem{scheres2012relion}
Sjors H.~W. Scheres.
\newblock {RELION}: Implementation of a {B}ayesian approach to cryo-{EM} structure determination.
\newblock {\em J. Struct. Biol.}, 180(3):519--530, 2012.

\bibitem{punjani2017cryosparc}
Ali Punjani, John~L. Rubinstein, David~J. Fleet, and Marcus~A. Brubaker.
\newblock {cryoSPARC}: Algorithms for rapid unsupervised cryo-{EM} structure determination.
\newblock {\em Nat. Methods}, 14(3):290--296, 2017.

\bibitem{dempster1977maximum}
Arthur~P. Dempster, Nan~M. Laird, and Donald~B. Rubin.
\newblock Maximum likelihood from incomplete data via the {EM} algorithm.
\newblock {\em J. R. Stat. Soc. Ser. B}, 39(1):1--38, 1977.

\bibitem{10.1042/BCJ20210708}
Dari Kimanius, Liyi Dong, Grigory Sharov, Takanori Nakane, and Sjors H.~W. Scheres.
\newblock New tools for automated cryo-{EM} single-particle analysis in {RELION}-4.0.
\newblock {\em Biochem. J.}, 478(24):4169--4185, 12 2021.

\bibitem{VANHEEL1987111}
Marin {Van Heel}.
\newblock Angular reconstitution: A posteriori assignment of projection directions for {3D} reconstruction.
\newblock {\em Ultramicroscopy}, 21(2):111--123, 1987.

\bibitem{Vainshtein1986DeterminationOT}
Boris~K. Vainshtein and Alexander~B. Goncharov.
\newblock Determination of the spatial orientation of arbitrarily arranged identical particles of unknown structure from their projections.
\newblock {\em Proc. llth Intern. Congr. on Elec. Micro.}, pages 459--460, 1986.

\bibitem{Bandeira_2020}
Afonso~S Bandeira, Yutong Chen, Roy~R Lederman, and Amit Singer.
\newblock Non-unique games over compact groups and orientation estimation in cryo-{EM}.
\newblock {\em Inverse Problems}, 36(6):064002, apr 2020.

\bibitem{singer2010detecting}
Amit Singer, Ronald~R. Coifman, Fred~J. Sigworth, David~W. Chester, and Yoel Shkolnisky.
\newblock Detecting consistent common lines in cryo-{EM} by voting.
\newblock {\em J. Struct. Biol.}, 169(3):312--322, 2010.

\bibitem{singer2011three}
Amit Singer and Yoel Shkolnisky.
\newblock Three-dimensional structure determination from common lines in cryo-{EM} by eigenvectors and semidefinite programming.
\newblock {\em SIAM J. Imaging Sci.}, 4(2):543--572, 2011.

\bibitem{shkolnisky2012viewing}
Yoel Shkolnisky and Amit Singer.
\newblock Viewing direction estimation in cryo-{EM} using synchronization.
\newblock {\em SIAM J. Imaging Sci.}, 5(3):1088--1110, 2012.

\bibitem{GREENBERG2017106}
Ido Greenberg and Yoel Shkolnisky.
\newblock Common lines modeling for reference free ab-initio reconstruction in cryo-{EM}.
\newblock {\em J. Struct. Biol.}, 200(2):106--117, 2017.

\bibitem{ZHAO2014153}
Zhizhen Zhao and Amit Singer.
\newblock Rotationally invariant image representation for viewing direction classification in cryo-{EM}.
\newblock {\em J. Struct. Biol.}, 186(1):153--166, 2014.

\bibitem{kam1980reconstruction}
Zvi Kam.
\newblock The reconstruction of structure from electron micrographs of randomly oriented particles.
\newblock {\em J. Theor. Biol.}, 82:15--39, 1980.

\bibitem{sharon2020method}
Nir Sharon, Joe Kileel, Yuehaw Khoo, Boris Landa, and Amit Singer.
\newblock Method of moments for {3D} single particle ab initio modeling with non-uniform distribution of viewing angles.
\newblock {\em Inverse Problems}, 36(4):044003, 2020.

\bibitem{bendory2023autocorrelation}
Tamir Bendory, Yuehaw Khoo, Joe Kileel, Oscar Mickelin, and Amit Singer.
\newblock Autocorrelation analysis for cryo-{EM} with sparsity constraints: Improved sample complexity and projection-based algorithms.
\newblock {\em Proc. Natl. Acad. Sci. USA}, 120:e2216507120, 2023.

\bibitem{kileel2025two}
Joe Kileel, Oscar Mickelin, Amit Singer, and Sheng Xu.
\newblock Two datasets are better than one: Method of double moments for {3D} reconstruction in cryo-{EM}.
\newblock {\em Inverse Problems}, 42(7):075016, 2026.

\bibitem{balanov2026provable}
Amnon Balanov, Tamir Bendory, Dan Edidin, Josh Katz, Shay Kreymer, and Nir Sharon.
\newblock Provable method of moments for cryo-{EM}.
\newblock {\em arXiv:2610.17892}, 2026.

\bibitem{BANDEIRA2023236}
Afonso~S. Bandeira, Ben Blum-Smith, Joe Kileel, Jonathan Niles-Weed, Amelia Perry, and Alexander~S. Wein.
\newblock Estimation under group actions: Recovering orbits from invariants.
\newblock {\em Appl. Comput. Harmon. Anal.}, 66:236--319, 2023.

\bibitem{doi:10.1137/23M155685X}
Tamir Bendory and Dan Edidin.
\newblock The sample complexity of sparse multireference alignment and single-particle cryo-electron microscopy.
\newblock {\em SIAM J. Math. Data Sci.}, 6(2):254--282, 2024.

\bibitem{tan2017addressing}
Yong~Zi Tan, Philip~R. Baldwin, Joseph~H. Davis, James~R. Williamson, Clinton~S. Potter, Bridget Carragher, and Dmitry Lyumkis.
\newblock Addressing preferred specimen orientation in single-particle cryo-{EM} through tilting.
\newblock {\em Nat. Methods}, 14(8):793--796, 2017.

\bibitem{kirian2012structure}
Richard~A. Kirian.
\newblock Structure determination through correlated fluctuations in {X}-ray scattering.
\newblock {\em J. Phys. B: At. Mol. Opt. Phys.}, 45(22):223001, 2012.

\bibitem{donatelli2015iterative}
Jeffrey~J. Donatelli, Peter~H. Zwart, and James~A. Sethian.
\newblock Iterative phasing for fluctuation {X}-ray scattering.
\newblock {\em Proc. Natl. Acad. Sci. USA}, 112(33):10286--10291, 2015.

\bibitem{kurta2017correlations}
Ruslan~P. Kurta, Jeffrey~J. Donatelli, Chun~Hong Yoon, et~al.
\newblock Correlations in scattered {X}-ray laser pulses reveal nanoscale structural features of viruses.
\newblock {\em Phys. Rev. Lett.}, 119(15):158102, 2017.

\bibitem{pande2018ab}
Kanupriya Pande, Jeffrey~J. Donatelli, Erik Malmerberg, Lutz Foucar, Christoph Bostedt, Ilme Schlichting, and Petrus~H. Zwart.
\newblock Ab initio structure determination from experimental fluctuation {X}-ray scattering data.
\newblock {\em Proc. Natl. Acad. Sci. USA}, 115(46):11772--11777, 2018.

\bibitem{zhao2024structure}
Wenyang Zhao, Osamu Miyashita, Miki Nakano, and Florence Tama.
\newblock Structure determination using high-order spatial correlations in single-particle {X}-ray scattering.
\newblock {\em IUCrJ}, 11(1):92--108, 2024.

\bibitem{bhamre2015orthogonal}
Tejal Bhamre, Teng Zhang, and Amit Singer.
\newblock Orthogonal matrix retrieval in cryo-electron microscopy.
\newblock In {\em 2015 IEEE 12th International Symposium on Biomedical Imaging (ISBI)}, pages 1048--1052, 2015.

\bibitem{Bhamre2017AnisotropicTF}
Tejal Bhamre, Teng Zhang, and Amit Singer.
\newblock Anisotropic twicing for single particle reconstruction using autocorrelation analysis.
\newblock {\em arXiv preprint arXiv:1704.07969}, 2017.

\bibitem{levin20183d}
Eitan Levin, Tamir Bendory, Nicolas Boumal, Joe Kileel, and Amit Singer.
\newblock {3D} ab initio modeling in cryo-{EM} by autocorrelation analysis.
\newblock In {\em Proc. IEEE Int. Symp. Biomed. Imaging (ISBI)}, pages 1569--1573, 2018.

\bibitem{doi:10.1137/22M1498218}
Shuai Huang, Mona Zehni, Ivan Dokmani{\'c}, and Zhizhen Zhao.
\newblock Orthogonal matrix retrieval with spatial consensus for {3D} unknown view tomography.
\newblock {\em SIAM J. Imaging Sci.}, 16(3):1398--1439, 2023.

\bibitem{doi:10.1137/24M1699644}
Jeremy Hoskins, Yuehaw Khoo, Oscar Mickelin, Amit Singer, and Yuguan Wang.
\newblock Subspace method of moments for ab initio {3-D} single particle cryo-{EM} reconstruction.
\newblock {\em SIAM J. Imaging Sci.}, 19(1):364--409, 2026.

\bibitem{andy2026overcrowded}
Andy Zhang, Oscar Mickelin, and Amit Singer.
\newblock Image moment estimation for crowded datasets.
\newblock {\em arXiv:2610.22840}, 2026.

\bibitem{bhamre2016denoising}
Tejal Bhamre, Teng Zhang, and Amit Singer.
\newblock Denoising and covariance estimation of single particle cryo-{EM} images.
\newblock {\em J. Struct. Biol.}, 195(1):72--81, 2016.

\bibitem{shi2022contrast}
Yunpeng Shi and Amit Singer.
\newblock Ab-initio contrast estimation and denoising of {cryo-EM} images.
\newblock {\em Comput. Methods Programs Biomed.}, 224:107018, 2022.

\bibitem{ROHOU2015216}
Alexis Rohou and Nikolaus Grigorieff.
\newblock {CTFFIND4}: Fast and accurate defocus estimation from electron micrographs.
\newblock {\em J. Struct. Biol.}, 192(2):216--221, 2015.

\bibitem{abramowitz1964handbook}
Milton Abramowitz and Irene~A. Stegun.
\newblock {\em Handbook of Mathematical Functions with Formulas, Graphs, and Mathematical Tables}, volume~55 of {\em National Bureau of Standards Applied Mathematics Series}.
\newblock U.S. Government Printing Office, Washington, DC, 1964.

\bibitem{Meyer_1954}
Burnett Meyer.
\newblock On the symmetries of spherical harmonics.
\newblock {\em Canad. J. Math.}, 6:135–157, 1954.

\bibitem{chirikjian2016harmonic}
Gregory~S. Chirikjian and Alexander~B. Kyatkin.
\newblock {\em Harmonic Analysis for Engineers and Applied Scientists: Updated and Expanded Edition}.
\newblock Dover Publications, Mineola, NY, 2016.

\bibitem{scheiderer2006sums}
Claus Scheiderer.
\newblock Sums of squares on real algebraic surfaces.
\newblock {\em Manuscripta Math.}, 119(4):395--410, 2006.

\bibitem{marshall2023fast}
Nicholas~F. Marshall, Oscar Mickelin, Yunpeng Shi, and Amit Singer.
\newblock Fast principal component analysis for cryo-electron microscopy images.
\newblock {\em Biol. Imaging}, 3:e2, 2023.

\bibitem{Zhao2014FastSP}
Zhizhen Zhao, Yoel Shkolnisky, and Amit Singer.
\newblock Fast steerable principal component analysis.
\newblock {\em IEEE Trans. Comput. Imaging}, 2(1):1--12, 2016.

\bibitem{zhao2013fourier}
Zhizhen Zhao and Amit Singer.
\newblock {F}ourier--{B}essel rotational invariant eigenimages.
\newblock {\em J. Opt. Soc. Am. A}, 30(5):871--877, 2013.

\bibitem{marshall2023fle}
Nicholas~F. Marshall, Oscar Mickelin, and Amit Singer.
\newblock Fast expansion into harmonics on the disk: A steerable basis with fast radial convolutions.
\newblock {\em SIAM J. Sci. Comput.}, 45:A2431--A2457, 2023.

\bibitem{nguyen2016trisnrnp}
Thi Hoang~Duong Nguyen, Wojciech~P. Galej, Xiao-chen Bai, Chris Oubridge, Andrew~J. Newman, Sjors H.~W. Scheres, and Kiyoshi Nagai.
\newblock Cryo-{EM} structure of the yeast {U4/U6.U5} tri-{snRNP} at 3.7 {\aa} resolution.
\newblock {\em Nature}, 530(7590):298--302, 2016.

\bibitem{graf2009sampling}
Manuel Gr{\"a}f and Daniel Potts.
\newblock Sampling sets and quadrature formulae on the rotation group.
\newblock {\em Numer. Funct. Anal. Optim.}, 30(7--8):665--688, 2009.

\bibitem{graf2013efficient}
Manuel Gr{\"a}f.
\newblock {\em Efficient algorithms for the computation of optimal quadrature points on {Riemannian} manifolds}.
\newblock PhD thesis, Chemnitz University of Technology, 2013.

\bibitem{katsevich2023likelihood}
Anya Katsevich and Afonso~S. Bandeira.
\newblock Likelihood maximization and moment matching in low {SNR} {G}aussian mixture models.
\newblock {\em Comm. Pure Appl. Math.}, 76(4):788--842, 2023.

\bibitem{Fan2024Maximum}
Zhou Fan, Roy~R. Lederman, Yi~Sun, Tianhao Wang, and Sheng Xu.
\newblock Maximum likelihood for high-noise group orbit estimation and single-particle cryo-{EM}.
\newblock {\em Ann. Statist.}, 52(1):52--77, 2024.

\bibitem{byrd1995limited}
Richard~H. Byrd, Peihuang Lu, Jorge Nocedal, and Ciyou Zhu.
\newblock A limited memory algorithm for bound constrained optimization.
\newblock {\em SIAM J. Sci. Comput.}, 16(5):1190--1208, 1995.

\bibitem{SCHERES2015114}
Sjors H.~W. Scheres.
\newblock Semi-automated selection of cryo-{EM} particles in {RELION}-1.3.
\newblock {\em J. Struct. Biol.}, 189(2):114--122, 2015.

\bibitem{10.7554/eLife.03080}
Wilson Wong, Xiao-chen Bai, Alan Brown, Israel~S Fernandez, Eric Hanssen, Melanie Condron, Yan~Hong Tan, Jake Baum, and Sjors~HW Scheres.
\newblock Cryo-{EM} structure of the \textit{Plasmodium falciparum} 80{S} ribosome bound to the anti-protozoan drug emetine.
\newblock {\em eLife}, 3:e03080, jun 2014.

\bibitem{lee2017hcn1}
Chia-Hsueh Lee and Roderick MacKinnon.
\newblock Structures of the human {HCN1} hyperpolarization-activated channel.
\newblock {\em Cell}, 168(1-2):111--120, 2017.

\bibitem{10.1093/nar/gkac283}
Sung-Hui Yi, Valentyn Petrychenko, Jan~Erik Schliep, Akanksha Goyal, Andreas Linden, Ashwin Chari, Henning Urlaub, Holger Stark, Marina V Rodnina, Sarah Adio, and Niels Fischer.
\newblock Conformational rearrangements upon start codon recognition in human 48{S} translation initiation complex.
\newblock {\em Nucleic Acids Res.}, 50(9):5282--5298, 05 2022.

\bibitem{bacic2021alc1}
Luka Bacic, Guillaume Gaullier, Anton Sabantsev, Laura~C. Lehmann, Klaus Brackmann, Despoina Dimakou, Mario Halic, Graeme Hewitt, Simon~J. Boulton, and Sebastian Deindl.
\newblock Structure and dynamics of the chromatin remodeler {ALC1} bound to a {PARylated} nucleosome.
\newblock {\em eLife}, 10:e71420, 2021.

\bibitem{Dhakal2023}
Ashwin Dhakal, Rajan Gyawali, Liguo Wang, and Jianlin Cheng.
\newblock A large expert-curated cryo-{EM} image dataset for machine learning protein particle picking.
\newblock {\em Scientific Data}, 10(1):392, 2023.

\bibitem{https://doi.org/10.1002/pro.4792}
Elaine~C. Meng, Thomas~D. Goddard, Eric~F. Pettersen, Greg~S. Couch, Zach~J. Pearson, John~H. Morris, and Thomas~E. Ferrin.
\newblock {UCSF} {ChimeraX}: Tools for structure building and analysis.
\newblock {\em Protein Sci.}, 32(11):e4792, 2023.

\bibitem{doi:10.1073/pnas.2419140122}
Marc~Aurèle Gilles and Amit Singer.
\newblock Cryo-{EM} heterogeneity analysis using regularized covariance estimation and kernel regression.
\newblock {\em Proc. Natl. Acad. Sci. U.S.A.}, 122(9):e2419140122, 2025.

\bibitem{doi:10.1073/pnas.1919114117}
Yimo Han, Xiao Fan, Haozhe Wang, Fang Zhao, Christopher~G. Tully, Jing Kong, Nan Yao, and Nieng Yan.
\newblock High-yield monolayer graphene grids for near-atomic resolution cryoelectron microscopy.
\newblock {\em Proc. Natl. Acad. Sci. U.S.A.}, 117(2):1009--1014, 2020.

\bibitem{scheres2009maximum}
Sjors H.~W. Scheres, Mikel Valle, Patricia Grob, Eva Nogales, and Jos{\'e}-Mar{\'\i}a Carazo.
\newblock Maximum likelihood refinement of electron microscopy data with normalization errors.
\newblock {\em J. Struct. Biol.}, 166(2):234--240, 2009.

\bibitem{zeng2007maximum}
Xiangyan Zeng, Henning Stahlberg, and Nikolaus Grigorieff.
\newblock A maximum likelihood approach to two-dimensional crystals.
\newblock {\em J. Struct. Biol.}, 160(3):362--374, 2007.

\bibitem{mcmullan2015thon}
G.~McMullan, K.~R. Vinothkumar, and R.~Henderson.
\newblock {T}hon rings from amorphous ice and implications of beam-induced {B}rownian motion in single particle electron cryo-microscopy.
\newblock {\em Ultramicroscopy}, 158:26--32, 2015.

\bibitem{scheres2014beam}
Sjors H.~W. Scheres.
\newblock Beam-induced motion correction for sub-megadalton cryo-{EM} particles.
\newblock {\em eLife}, 3:e03665, 2014.

\bibitem{zivanov2019bayesian}
Jasenko Zivanov, Takanori Nakane, and Sjors H.~W. Scheres.
\newblock A {B}ayesian approach to beam-induced motion correction in cryo-{EM} single-particle analysis.
\newblock {\em IUCrJ}, 6(1):5--17, 2019.

\bibitem{li2013electron}
Xueming Li, Paul Mooney, Shawn Zheng, Christopher~R. Booth, Michael~B. Braunfeld, Sander Gubbens, David~A. Agard, and Yifan Cheng.
\newblock Electron counting and beam-induced motion correction enable near-atomic-resolution single-particle cryo-{EM}.
\newblock {\em Nat. Methods}, 10(6):584--590, 2013.

\bibitem{zheng2017motioncor2}
Shawn~Q. Zheng, Eugene Palovcak, Jean-Paul Armache, Kliment~A. Verba, Yifan Cheng, and David~A. Agard.
\newblock {MotionCor2}: Anisotropic correction of beam-induced motion for improved cryo-electron microscopy.
\newblock {\em Nat. Methods}, 14(4):331--332, 2017.

\bibitem{bendory2023toward}
Tamir Bendory, Nicolas Boumal, William Leeb, Eitan Levin, and Amit Singer.
\newblock Toward single particle reconstruction without particle picking: Breaking the detection limit.
\newblock {\em SIAM J. Imaging Sci.}, 16(2):886--910, 2023.

\end{thebibliography}

\appendix

\section{Radial functions for the moments}
\label{app:radial}
By the Fourier-slice theorem and~\eqref{eqn:sfb3d}, the radial part of any projection of $\Phi_A$ at angular frequency $q$ is
\begin{equation}
    \sum_{l=|q|}^{L}\sum_{s=1}^{S_l}\sum_{m=-l}^{l} A_{l,m,s}\,D^l_{m,q}(R)\,F_{l,s}(r)\,Y_l^{q}(\tfrac{\pi}{2},0),
\end{equation}
a linear combination of the slices $F_{l,s}(r)\,Y_l^{q}(\tfrac{\pi}{2},0)$ with $l\equiv q \pmod 2$ and $|q|\le l\le L$. Components of the images outside the span of these slices cannot be produced by any $(A,B)$. We therefore expand the images in an orthonormal basis of this span. Since the slices are nearly linearly dependent, we compute their singular value decomposition on a radial quadrature grid, and take $f_{q,1},\dots,f_{q,T_q}$ to be the left singular vectors whose singular values exceed $1\%$ of the largest one. 

\section{Moment representations in a steerable basis}
\label{app:moment_derivation}

We derive the representations of moments~\eqref{eqn:m1quad}--\eqref{eqn:m3quad} in this section. By Section~\ref{sec:kam}, it suffices to consider the entries with $q=0$ in $M_1$, $q_1=q_2$ in $M_2$, and $q_1+q_2+q_3=0$ in $M_3$, since the others vanish. We first insert~\eqref{eqn:coef_shift} into~\eqref{eqn:M1_population}--\eqref{eqn:M3_population} and average over the uniform angle $\psi$. Since $\frac{1}{2\pi}\int_0^{2\pi}e^{-in\psi}\,d\psi=\delta_{n,0}$, only the terms with $q'=0$ remain in $M_1$, those with $q_1'=q_2'$ in $M_2$, and those with $q_1'+q_2'+q_3'=0$ in $M_3$, and in these terms the factors $(-i)^{q-q'}$ multiply to one. It remains to integrate the products of $\mu_B(R)$ and the coefficients $a_{q',t'}(A,R,0)$ over $\SO(3)$.
We first recall the Wigner-D matrices~\cite{chirikjian2016harmonic}. For $R\in\SO(3)$ and $l\ge0$, the Wigner-D matrix $D^l(R)=(D^l_{m,m'}(R))_{m,m'=-l}^l$ describes how the spherical harmonics of degree $l$ change under rotation:
\begin{equation}
    Y_l^m(R^T\eta)=\sum_{m'=-l}^l D_{m,m'}^l(R)\, Y_l^{m'}(\eta),\qquad \eta\in\mathbb{S}^2.
\end{equation}
It also holds that
\begin{equation}
    D^l(R_1R_2)=D^l(R_2)\,D^l(R_1).
    \label{eqn:wigner-product}
\end{equation}

Let
\begin{equation}
    R_z(\gamma)=\begin{pmatrix}\cos\gamma&-\sin\gamma&0\\ \sin\gamma&\cos\gamma&0\\ 0&0&1\end{pmatrix}
    \label{eqn:rotz}
\end{equation}
and
\begin{equation}
    R_y(\theta)=\begin{pmatrix}\cos\theta&0&\sin\theta\\ 0&1&0\\ -\sin\theta&0&\cos\theta\end{pmatrix}
\end{equation}
be the elementary rotations about the $z$-axis and the $y$-axis.  We use the fact that every $R\in\SO(3)$ can be written as
\begin{equation}
    R=R_z(\gamma)R_y(\theta)R_z(\varphi)
\end{equation}
for some $\varphi\in[0,2\pi)$, $\theta\in[0,\pi]$ and $\gamma\in[0,2\pi)$, where $(\theta,\varphi)$ are the spherical coordinates of the viewing direction $Re_3$ and $\gamma$ is the in-plane rotation angle. The Haar measure can be written as 
\begin{equation}
    dR=\frac{1}{8\pi^2}\sin\theta\,d\varphi\,d\theta\,d\gamma, 
\end{equation}
so for any integrable function  $g:\SO(3) \to \C$, 
\begin{equation}
    \begin{aligned}
    \int_{\SO(3)}g(R)\,dR=&\frac{1}{8\pi^2}\int_0^{2\pi}\int_0^{\pi}\int_0^{2\pi}g\big(R_z(\gamma)R_y(\theta)R_z(\varphi)\big) d\gamma\,\sin\theta\,d\theta\,d\varphi .
    \end{aligned}
    \label{eqn:haar}
\end{equation}

For a point $\eta\in S^2$ with spherical coordinates
$(\vartheta,\varphi)$, the point $R_z(\gamma)\eta$ has
coordinates $(\vartheta,\varphi+\gamma)$. Therefore, we have 
\begin{equation}
    \begin{split}
        Y_l^m(R_z(\gamma)\eta)=e^{im\gamma}Y_l^m(\eta), \\ 
        D^l_{m,m'}(R_z(\gamma))=\delta_{m,m'}e^{-im\gamma}.
    \end{split}
\end{equation}
Together with~\eqref{eqn:wigner-product}, they imply
\begin{equation}
    D^l_{m,q}(R_z(\gamma)R)=D^l_{m,q}(R)\,e^{-iq\gamma}.
    \label{eqn:inplane_rotate}
\end{equation}
By~\eqref{eqn:coef2d} and~\eqref{eqn:inplane_rotate}, rotating the projection in-plane only changes the phase of its coefficients,
\begin{equation}
    a_{q,t}(A,R_z(\gamma)R,0)=e^{-iq\gamma}\,a_{q,t}(A,R,0).
    \label{eqn:phase}
\end{equation}

For $M_3$, we write such an integrand as
\begin{equation}
    g(R)=\mu_B(R)\prod_{i=1}^{3}a_{q_i',t_i'}(A,R,0)
\end{equation}
with $q_1'+q_2'+q_3'=0$. By~\eqref{eqn:phase} and the invariance of $\mu_B$, $g$ does not depend on the in-plane rotation,
\begin{equation}
g\big(R_z(\gamma)R_y(\theta)R_z(\varphi)\big)=g\big(R_y(\theta)R_z(\varphi)\big).
\end{equation}
Inserting this into~\eqref{eqn:haar} gives
\begin{equation}
    \begin{aligned}
    \int_{\SO(3)}g(R)\,dR=\frac{1}{4\pi}\int_0^{2\pi}\int_0^{\pi}g\big(R_y(\theta)R_z(\varphi)\big)\sin\theta\,d\theta\,d\varphi .
    \end{aligned}
    \label{eqn:m3sphere}
\end{equation}
The formulas for $M_1$ and $M_2$ can be derived in a similar way.

The integrand in~\eqref{eqn:m3sphere} is a function on $\mathbb{S}^2$ of the viewing direction $(\theta,\varphi)$. We now bound its degree as a spherical harmonic series. Setting $\eta=e_3$ in the definition of $D^l$ and using $Y_l^{m'}(e_3)=\sqrt{(2l+1)/(4\pi)}\,\delta_{m',0}$, we get
\begin{equation}
    D^l_{m,0}(R)=\sqrt{\frac{4\pi}{2l+1}}\,Y_l^m(R^Te_3).
\end{equation}
In other words, up to constants, the Wigner-D entries with second index $0$ are the spherical harmonics of the viewing direction $R^Te_3$. By~\eqref{eqn:coef2d}, each coefficient $a_{q,t}(A,R,0)$ is a linear combination of the Wigner-D entries $D^l_{m,q}(R)$ with $l\le L$. Since each $p_B^{(i)}$ in~\eqref{eqn:viewing_sos} has degree at most $P/2$, $\mu_B$ is a spherical harmonic series of degree at most $P$, i.e., a linear combination of $D^l_{m,0}(R)$ with $l\le P$. The product can be expanded as
\begin{equation}
    D^{l_1}_{m_1,q_1}D^{l_2}_{m_2,q_2}=\sum_{l=|l_1-l_2|}^{l_1+l_2}c_l\,D^{l}_{m_1+m_2,q_1+q_2}
\end{equation}
where $c_l$ are the Clebsch--Gordan coefficients~\cite{chirikjian2016harmonic}. Hence, $g$, as a product of three coefficients and $\mu_B$, is a linear combination of $D^l_{m,q_1'+q_2'+q_3'}(R)=D^l_{m,0}(R)$ with $l\le 3L+P$, which is a spherical harmonic series of degree at most $3L+P$. Similarly, the integrands of $M_1$ and $M_2$ have degree at most $L+P$ and $2L+P$. We can therefore construct a quadrature rule with nodes $(\theta_k,\varphi_k)$ and weights $w_k$, $k=1,\dots,K$, that integrates spherical harmonic series of degree at most $3L+P$ exactly~\cite{graf2009sampling,graf2013efficient}, so that
\begin{equation}
    \begin{aligned}
    &\frac{1}{4\pi}\int_0^{2\pi}\int_0^{\pi}g\big(R_y(\theta)R_z(\varphi)\big)\sin\theta\,d\theta\,d\varphi=\sum_{k=1}^K w_k\,g(\widetilde R_k),\qquad \widetilde R_k=R_y(\theta_k)R_z(\varphi_k).
    \end{aligned}
\end{equation}
Applying the quadrature formulas gives~\eqref{eqn:m1quad},~\eqref{eqn:m2quad} and~\eqref{eqn:m3quad}.

\end{document}